\documentclass[a4paper,english,aps,reprint]{revtex4-2}
\usepackage[T1]{fontenc}
\usepackage[utf8]{inputenc}
\usepackage{babel}
\usepackage{amsmath}
\usepackage{amssymb}
\usepackage{stackrel}
\usepackage{graphicx}
\usepackage{esint}
\usepackage[unicode=true,pdfusetitle,
 bookmarks=true,bookmarksnumbered=false,bookmarksopen=true,bookmarksopenlevel=1,
 breaklinks=false,pdfborder={0 0 0},backref=false,colorlinks=false]
 {hyperref}

\usepackage{xcolor}
\definecolor{change}{RGB}{69, 139, 160}

\makeatletter

\pdfpageheight\paperheight
\pdfpagewidth\paperwidth

\@ifundefined{textcolor}{}
{%
 \definecolor{BLACK}{gray}{0}
 \definecolor{WHITE}{gray}{1}
 \definecolor{RED}{rgb}{1,0,0}
 \definecolor{GREEN}{rgb}{0,1,0}
 \definecolor{BLUE}{rgb}{0,0,1}
 \definecolor{CYAN}{cmyk}{1,0,0,0}
 \definecolor{MAGENTA}{cmyk}{0,1,0,0}
 \definecolor{YELLOW}{cmyk}{0,0,1,0}
 \definecolor{CHANGE}{rgb}{0.27, 0.545, 0.627}
}

\usepackage{babel}
\usepackage{babel}

\usepackage{babel}
\usepackage[normalem]{ulem}
\usepackage{babel}
\usepackage{babel}
\usepackage{amsfonts}
\providecommand{\U}[1]{\protect\rule{.1in}{.1in}}

\pdfpageheight\paperheight
\pdfpagewidth\paperwidth

\@ifundefined{textcolor}{}{
\definecolor{BLACK}{gray}{0}
\definecolor{WHITE}{gray}{1}
\definecolor{RED}{rgb}{1,0,0}
\definecolor{GREEN}{rgb}{0,1,0}
\definecolor{BLUE}{rgb}{0,0,1}
\definecolor{CYAN}{cmyk}{1,0,0,0}
\definecolor{MAGENTA}{cmyk}{0,1,0,0}
\definecolor{YELLOW}{cmyk}{0,0,1,0}
}

\makeatother

\begin{document}

\title{One-dimensional model potentials optimized for the calculation of the HHG spectrum}

\author{Krisztina Sallai\textsuperscript{1,2}, Szabolcs Hack\textsuperscript{1,2}, Szilárd Majorosi\textsuperscript{1}, and Attila Czirják\textsuperscript{1,2}}

\email{czirjak@physx.u-szeged.hu}

\affiliation{
 \textsuperscript{1}ELI-ALPS, ELI-HU Non-Profit Ltd., Wolfgang Sandner utca 3., Szeged, H-6728, Hungary
\linebreak{}
\textsuperscript{2}Department of Theoretical Physics, University
of Szeged, Tisza L. krt. 84-86., Szeged, H-6720, Hungary
}
\begin{abstract}
Based on the favourable properties of previously used one-dimensional (1D) atomic model potentials, we introduce a novel 1D atomic model potential for
the 1D simulation of the quantum dynamics of a single active electron atom driven by a strong, linearly polarized near-infrared laser pulse. By comparing numerical simulation results of typical strong-field physics scenarios in 1D and 3D, we show that this novel 1D model potential gives single atom HHG spectra with impressively increased accuracy for the most frequently used driving laser pulse parameters.

\end{abstract}
\maketitle

\section{Introduction\label{sec:introduction}}

Atomic and molecular physics has been revolutionised by the appearance of attosecond pulses \cite{hentschel2001attosecond,kienberger2002attosecond,drescher2002attosecond,baltuska2003attosecond,Uiberacker_Nature_2007,Krausz_RevModPhys_2009_Attosecond_physics,hommelhoff2009extremelocalization,Schultze_Science_2010,Haessler_NatPhys_2010,pfeiffer2012attoclock,shafir2012tunneling,ranitovic2014attosecondcontrol,peng2015attosecondtracing,ciappina2017attosecondnano}.
In order to fully understand the processes in attosecond and strong-field
physics, the quantum evolution of an involved atomic system driven by a strong laser pulse is often needed \cite{Keldysh_JETP_1965,varro1993multiphotonequation,lewenstein1994hhgtheory,protopapas1997tdseionization,ivanov2005strongfield,gordon2005tdsecoulomb,frolov2012attosecondanalytic}.
Nevertheless, a non-perturbative range is reached in these calculations, therefore the analytical  solution of the corresponding Schrödinger equation is unattainable, except for the simplest cases. Thus, it is very important and necessary to use suitable approximations, which may also make the numerical solutions more effective in certain cases.

In the case of linearly polarized driving laser pulses, the main dynamics happen along the direction of the 
electric field of the laser pulse, which underlies the success of some one-dimensional (1D) approximations \cite{eberly1988softcoulombspecra,eberly1991softcoulombatom,bauer1997tdse1dhemodel,chirilua2010HHGemission,silaev2010tdsecoulombs,sveshnikov2012schrodingercoulomb,graefe2012quantumphasespace,czirjak2000ionizationwigner,czirjak2013rescatterentanglement,geltman2011boundstatesdelta,baumann2015wignerionization,teeny2016ionizationtime}.
Various 1D atomic model potentials are in use to account for the behavior of the atomic system, but the chosen model potential may heavily influence the 1D results and their comparison with the true three-dimensional (3D) results is usually non-trivial. Regarding high-order harmonic generation (HHG), e.g. the induced dipole moment may largely deviate in the 1D simulation from that in the 3D simulation, if the very same laser electric field is applied. 
The 1D model potentials play an important role in discovering novel quantum features, for example for optimizing strong-field processes for different aims \cite{chomet2019bridge,chomet2021phase,chomet2022machine}, in exploring quantum entanglement \cite{majorosi2017entanglement,czirjak2013rescatterentanglement} and the details of tunneling \cite{hack2021interference} in strong-field ionization. Proper model potentials are crucial in ensemble simulations of strong-field processes based on classical \cite{sarkadi2020nonseq} or semi-classical dynamics \cite{truong2022soft}. 
The 1D model potentials were also successfully used recently for  strong-field ionization processes \cite{strelkov2023autoion,harris2023ati,kocak2020one}, and in simulations of atoms driven by XUV pulses \cite{ziems2023xuv,lan2023twocolor,wang2021sch,wang2022mom}. 

In the present paper, we introduce a novel 1D atomic model potential for atomic strong-field quantum dynamics driven by a linearly polarized near-infrared (NIR) laser pulse. In a previous work \cite{majorosi2018improved}, some of the present authors introduced a modified soft-core Coulomb potential, for which the ground state density is very close to the reduced 3D atomic ground state density. This modified soft-core Coulomb potential gives a very accurate high harmonic spectrum in the lower frequency range, but it usually overestimates the high frequency part by 1-2 orders of magnitude. Although this can be well corrected with a scaling function, we recently explored the underlying problem which then led us to a new 1D atomic model potential that gives the HHG spectrum more precisely both in the lower and in the  higher frequency range. Our key idea is to use the benefits of both the soft-core Coulomb potential and the modified soft-core Coulomb potential. 
In the present paper, we show how to combine these to obtain a novel 1D atomic model potential and then we present the results of applying it in numerical simulations of  typical strong-field physics scenarios. We use atomic units in this paper.

\section{3D and 1D model systems\label{sec:3D-and-1D-model-systems}}

\subsection{3D reference system and 1D model system\label{sec:3D-1D}}

For the solution of the 3D and 1D models, we use the same method as in \cite{majorosi2018improved}, see also \cite{majorosi2016tdsesolve}. 
Here, we recall the most important formulas. We consider the action of a linearly polarized laser pulse on the single active electron atom in dipole approximation and in length gauge:
\begin{equation}
V_{\text{ext}}(z,t)=z\cdot\mathcal{E}_{z}(t),\label{eq:pot1_length_z}
\end{equation}
and we seek solutions of the time-dependent Schrödinger equation 
\begin{equation}
i\frac{\partial}{\partial t}\Psi^{\mathrm{3D}}\left(z,\rho,t\right)=\left[H_{0}^{\mathrm{3D}}+V_{\text{ext}}(z,t)\right]\Psi^{\mathrm{3D}}\left(z,\rho,t\right)\label{eq:3d_tdse},
\end{equation}
with
\begin{equation}
H_{0}^{{\mathrm{3D}}}=T_{z}+T_{\rho}-\frac{Z}{\sqrt{\rho^{2}+z^{2}}},\label{eq:3d_hamiltonian_0}
\end{equation}
where $Z$ is the effective charge of the ion-core ($Z=1$ for hydrogen) and the two relevant terms of the kinetic energy operator are given by
\begin{equation}
T_{\rho}=-\frac{1}{2\mu}\left[\frac{\partial^{2}}{\partial\rho^{2}}+\frac{1}{\rho}\frac{\partial}{\partial\rho}\right],\,\,\,\,\,\,\,\,\,\, T_{z}=-\frac{1}{2\mu}\frac{\partial^{2}}{\partial z^{2}}.\label{eq:3d_kinetic_z_rho}
\end{equation}
In order to model the 3D time-dependent process in 1D, we use a 1D atomic Hamiltonian of the following form: 
\begin{equation}
H_{0}^{{\mathrm{1D}}}=T_{z}+V_{0}^{{\mathrm{1D}}}(z),\label{eq:1d_hamiltonian_0}
\end{equation}
where $V_{0}^{{\mathrm{1D}}}(z)$ is a 1D atomic model potential of choice, and then to seek solutions of the time-dependent Schrödinger equation
\begin{equation}
i\frac{\partial}{\partial t}\Psi^{{\mathrm{1D}}}\left(z,t\right)=\left[H_{0}^{{\mathrm{1D}}}+V_{\text{ext}}(z,t)\right]\Psi^{{\mathrm{1D}}}\left(z,t\right),\label{eq:1d_tdse}
\end{equation}
where the laser potential $V_{\text{ext}}(z,t)$ is given in \eqref{eq:pot1_length_z}. In this article we are going to introduce a new form of $V_{0}^{{\mathrm{1D}}}(z)$ to model the high harmonic spectrum as correctly as possible. But before doing so, let us shortly summarize the most important features of two of the 1D potentials used earlier. 

\subsection{
1D soft-core Coulomb model potentials\label{subsec:1D SC}}

There are a number of well-known 1D atomic model potentials in the
literature \cite{silaev2010tdsecoulombs,sveshnikov2012schrodingercoulomb,graefe2012quantumphasespace, majorosi2018improved},
having their advantages and disadvantages. Here we summarize the basics
of two of these, which we think to be the most important for the modeling
of strong-field phenomena. For convenience, we use $\mu=1$ for the electron's relative mass.

The soft-core Coulomb (SC) potential is defined as 
\begin{equation}
V_{{\rm SC}}^{{\rm 1D}}(z)=-\frac{Z}{\sqrt{z^{2}+\alpha^{2}}}\label{eq:pot1_sc}
\end{equation}
where the smoothing parameter $\alpha$ is usually adjusted to match
the ground state energy 
of a selected atom. 
If this is done in single active electron approximation, using an effective ion-core charge $Z$, then setting $\alpha = \sqrt{2}/Z$ gives the following ground state energy: 
\begin{equation}
 E_{{\rm SC}} =-\frac{Z^2}{2},
 \end{equation}
 and wave function \cite{liu1992scgroundstate}:
 \begin{equation}
 \psi_{{\rm SC}}(z) =\mathcal{N}_{{\rm SC}}\left(1+\sqrt{z^{2}+\alpha^2}\right)e^{-\sqrt{z^{2}+\alpha^2}}\label{eq:pot1_sc_groundsys}
\end{equation}
where $\mathcal{\mathcal{N}}_{{\rm SC}}$ is the normalization factor. The most important features of this model potential are that it is
a smooth function, and it has an asymptotic Coulomb form and Rydberg continuum. For hydrogen, $Z=1$ and the ground state energy is $-0.5$, the first excited state is at $E_{1,{\rm SC}}=-0.2329034$.

The improved soft-core Coulomb potential with modified parameters (MSC), introduced in \cite{majorosi2018improved}, is defined as
\begin{equation}
V_{{\rm MSC}}^{{\rm 1D}}(z)=-\frac{\frac{1}{2}Z}{\sqrt{z^{2}+\frac{1}{4Z^{2}}}},\label{eq:pot1_sc_corr}
\end{equation}
which also gives the desired ground state energy $E_{{\rm MSC}}=-Z^{2}/2$. For hydrogen, the energy of its first excited state is
$E_{1,{\rm MSC}}=-0.1058670$.
It has $\frac{1}{2}Z/|z|$ asymptotic behavior which ensures that the ground state density is very close to the reduced density of the 3D ground state. Using this potential leads to improved results in strong-field simulations, especially in the lower frequency regime.

\begin{figure*}
\begin{raggedright} \hspace{4.6cm}(a)\hspace{8.7cm}(b) 

\end{raggedright}

\includegraphics[width=1\columnwidth]{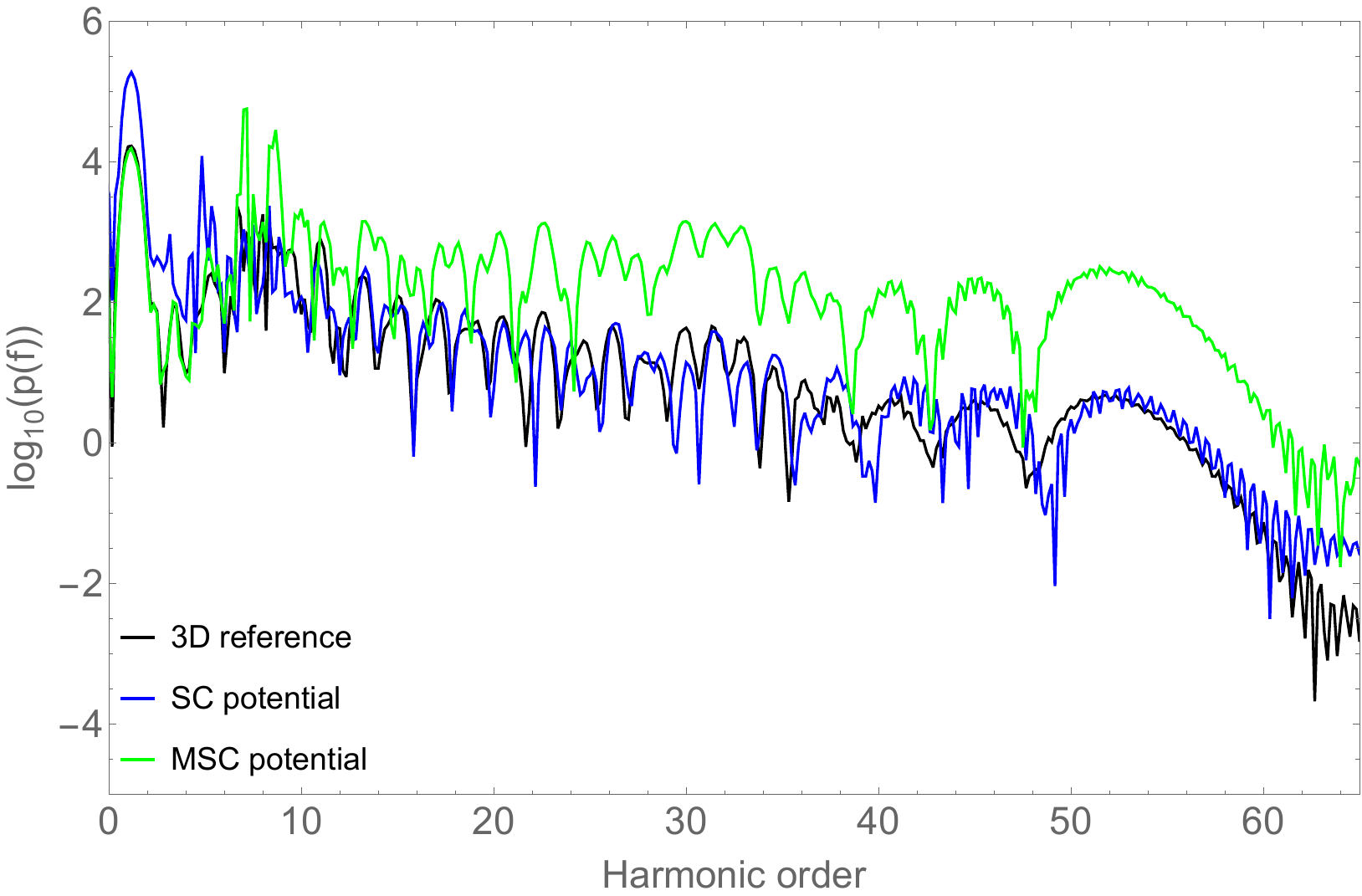}\hspace{0.5cm}\includegraphics[width=1\columnwidth]{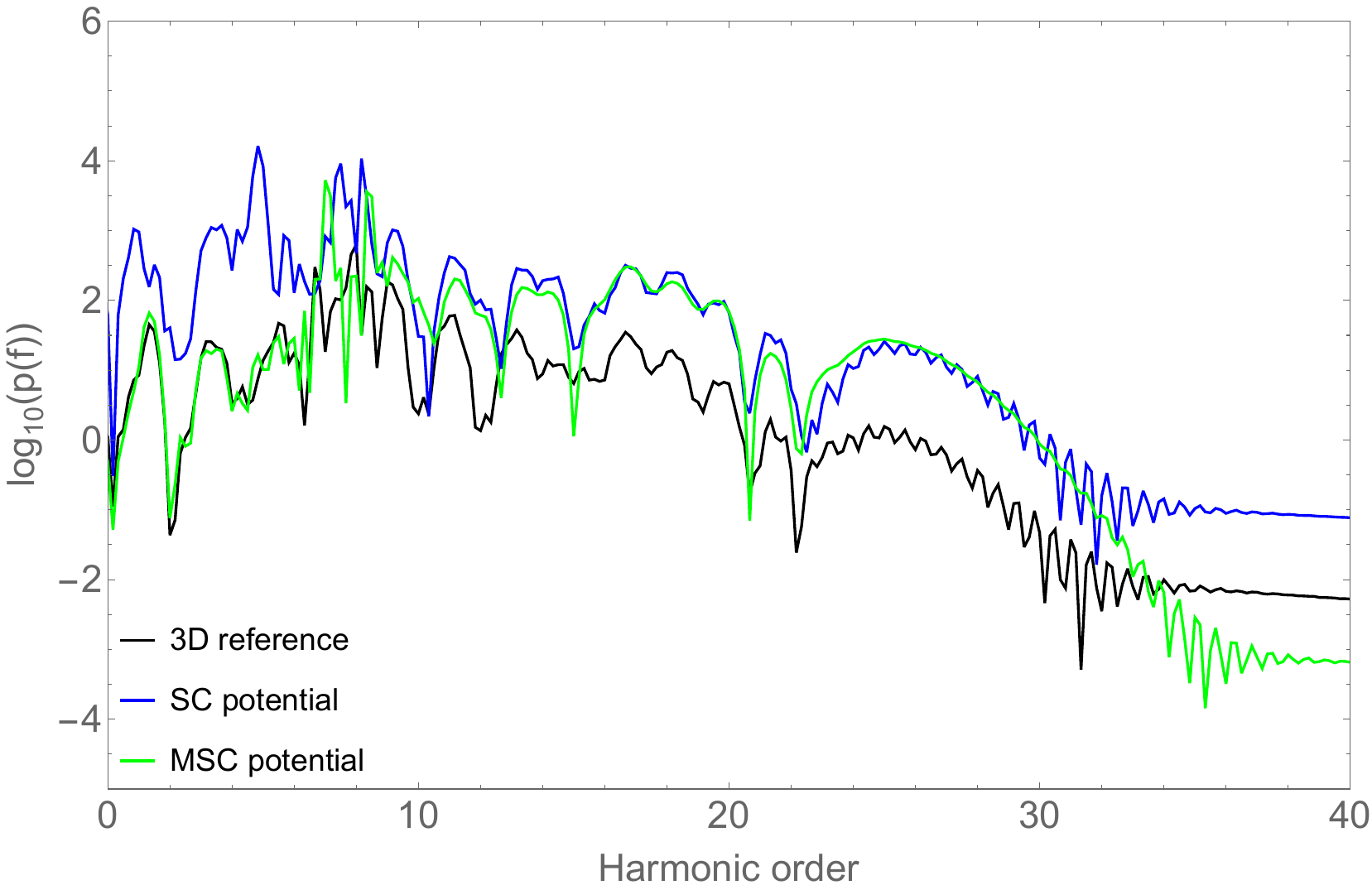}

\protect\protect\caption{The HHG spectrum of a hydrogen atom obtained with the SC potential (blue) and the MSC  potential (green), driven by a laser pulse with parameters $N_{\text{C}}=3$, $T=110$ and (a) $F=0.1$, (b) $F=0.06$. Results of the corresponding 3D simulation are plotted in black.}

\label{fig:T110_HHG_SC_MSC} 
\end{figure*}

\begin{figure*}
\begin{raggedright} 

\end{raggedright}

\includegraphics[width=1\columnwidth]{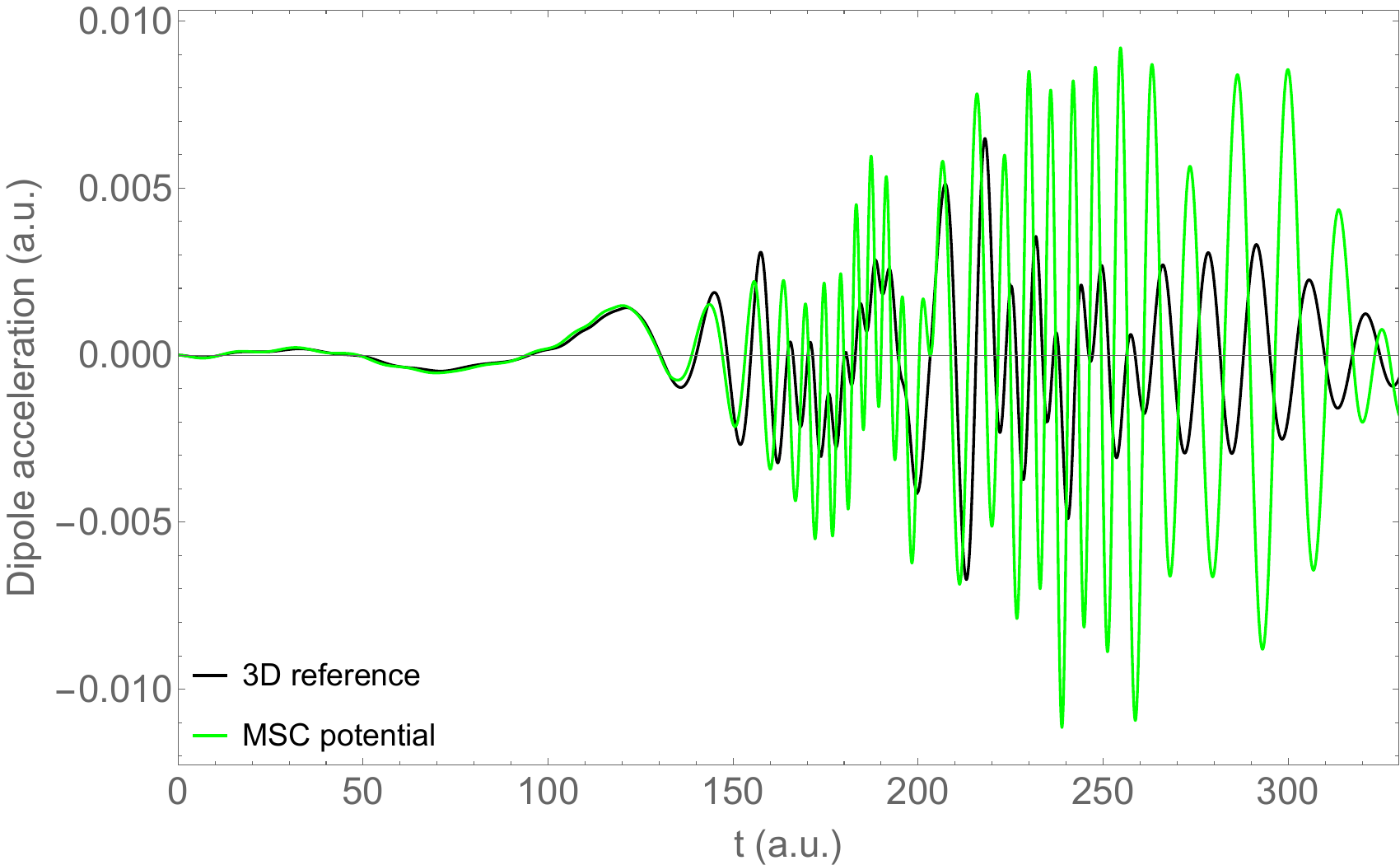}

\protect\protect\caption{The dipole acceleration for hydrogen atom, driven by a laser pulse with parameters $T=110$, $N_{\text{C}}=3$, $F=0.1$, as simulated by the MSC model potential (green) in comparison with the 3D reference (black).}

\label{fig:dipacc} 
\end{figure*}

Regarding the SC model potential \eqref{eq:pot1_sc}, experience shows that it does not give strong-field simulation results that would be quantitatively comparable to those of the reference 3D system (cf. \cite{bandrauk2009tdsehydrogen,graefe2012quantumphasespace,czirjak2012entanglementbuild})
therefore the model system parameters need to be manually adjusted, for example by changing the strength of $V_{\text{ext}}(z,t)$. In case of the HHG spectrum, using the SC model potential generally causes the HHG spectrum to be overestimated for all frequencies, but for certain special cases, the higher frequency response matches well with the 3D results (see Fig. \ref{fig:T110_HHG_SC_MSC}). With the MSC model potential \eqref{eq:pot1_sc_corr}, there is no need for changing $V_{\text{ext}}(z,t)$, however, when examining the HHG spectrum, the low frequencies are matched quite well, but it overestimates the spectrum for the higher frequencies (see Fig.  \ref{fig:T110_HHG_SC_MSC}). 
We found that the reason for the latter
is mainly the too large amplitude of high-frequency oscillations at the initial stage of the electron's escape (see Fig. \ref{fig:dipacc}.).

\section{
Definition of the novel 1D model potential\label{sec:pot1_model}}

\begin{figure*}
\begin{raggedright} 

\end{raggedright}

\includegraphics[width=1\columnwidth]{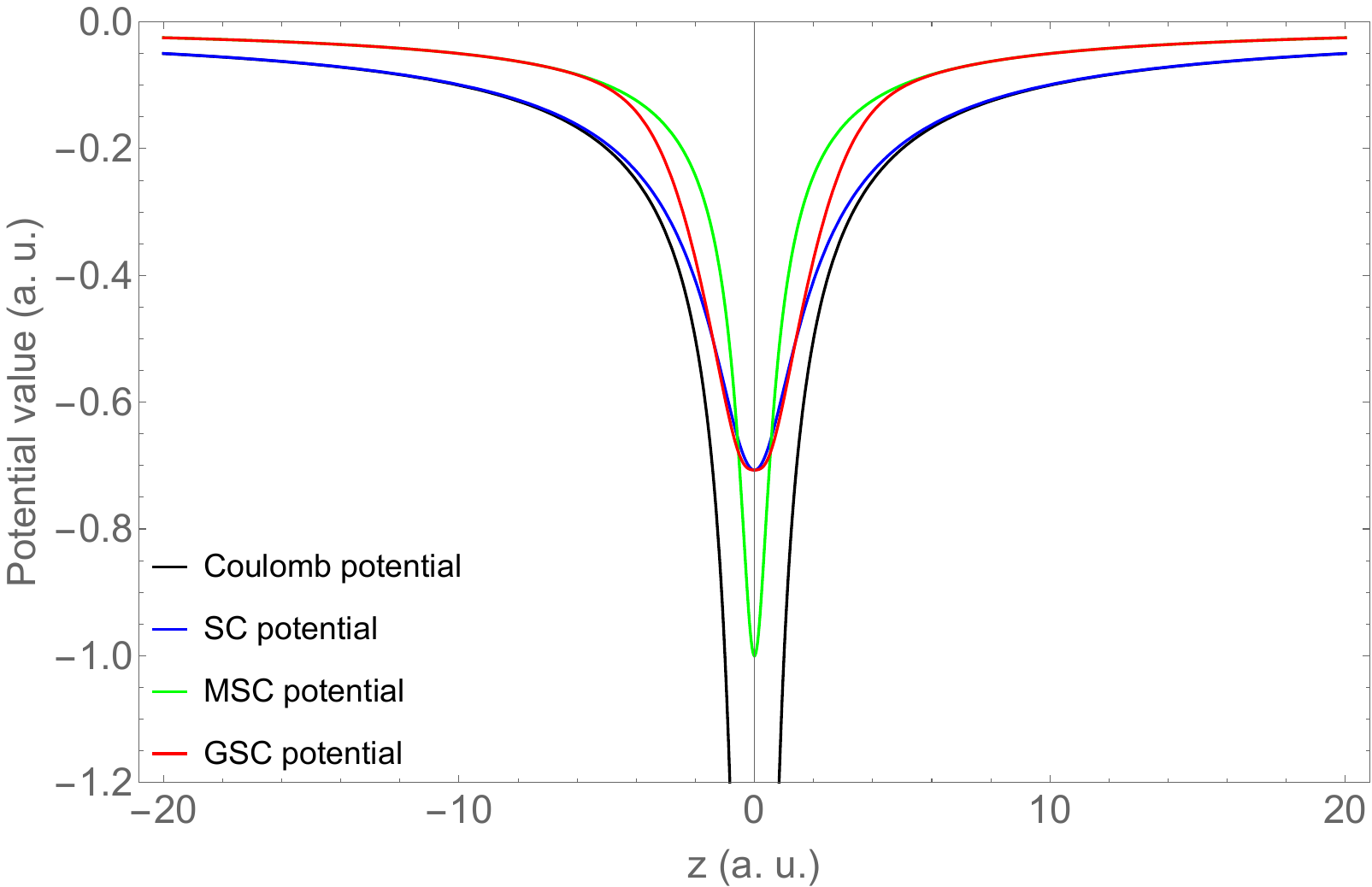}

\protect\protect\caption{Comparison of the 1D model potentials and the corresponding cut of the 3D Coulomb potential for hydrogen.}

\label{fig:potentials} 
\end{figure*}

We are now going to introduce a new formula for $V_{0}^{{\rm 1D}}(z)$. Our key idea leading to this novel 1D model potential is based on the fact that 
for the MSC potential \eqref{eq:pot1_sc_corr}, the lower frequency part of the HHG spectrum matches really well to the 3D reference, but it overestimates the power spectrum by around two orders of magnitude for the higher frequencies. With numerical testing we found that this is caused by the fact that the potential binds too strongly at the centre, see Fig. \ref{fig:potentials}.
(Note that even though it is commonly thought that the characteristics of the HHG spectrum are dominantly shaped by the electron wave packet which leaves the atom, the wave function around the center still has a significant effect on them.) 
Thus, the SC potential around the center seems to be more advantageous, since its HHG spectrum may better match with the 3D reference for higher frequencies, for certain parameters.
From these observations we constructed a novel 1D model potential which utilizes the favourable properties of both model potentials by joining them smoothly at an appropriate point around the centre: namely, it binds as the SC at the center, i.e. less strongly than the MSC, and it goes to zero like the MSC far away from the centre, i.e. faster than the SC. We chose well-behaving Gaussian window functions for joining the two constituent potentials. Thus, we define the novel 1D atomic model potential: the Gaussian windowed soft-core Coulomb (GSC) potential as:
\begin{equation}
\begin{split}
V^{\mathrm{1D}}_{\mathrm{GSC}}(z) = & V^{\mathrm{1D}}_{\mathrm{SC}}(z)\cdot e^{-(z/a)^2} \\ &+V^{\mathrm{1D}}_{\mathrm{MSC}}(z)\left(1 - e^{-(z/b)^2}\right),\label{eq:pot1_gsc}
\end{split}
\end{equation}
where with the parameters $a$ and $b$, we can ensure that the ground state energy is appropriate for the atom we aim to model. For the cases of hydrogen and argon, we found that the numerical values of $a=2.551$ and $b=2$ in \eqref{eq:pot1_gsc} give the proper ground state energy of the real 3D H and Ar atom. 

\section{Strong-field simulation results \label{sec:results}}

In this section, we present and compare the results of strong-field
simulations, obtained by solving \eqref{eq:1d_tdse} with the various 1D model potentials discussed in the previous sections and we compare the results with the solution of \eqref{eq:3d_tdse} as a 3D reference. We characterize the dynamics with the mean value of the dipole moment $\left\langle z\right\rangle (t)$, the dipole power spectrum $p(f)$ and the ground state population loss $g(t)$. The dipole power spectra $p(f)$ is one of the most important quantities for high order harmonic generation \cite{McPherson_JOSAB_1987_HHG,Harris_OptCom_1993_HHG_Atto,lewenstein1994hhgtheory}
and attosecond pulses. For the formulas of these physical quantities, see Appendix \ref{sec:comparable-physical-quantities} and for the numerical methods of the solution and some details about the numerical accuracy of the simulations, see \cite{majorosi2018improved}.

In these simulations, we model the linearly polarized few-cycle laser
pulse with a sine-squared envelope function. The corresponding time-dependent
electric field has non-zero values only in the interval $0\leq t\leq N_{\text{C}}T$
according to the formula:
\begin{equation}
\mathcal{E}_{z}(t)= F\cdot\sin^{2}\left(\frac{2\pi t}{2N_{\text{C}}T}\right)\cdot\cos\left(\left(\frac{2\pi}{T}+k\cdot t\right)\cdot t\right),\label{eq:sim_sinpulse3_E}
\end{equation}
where $T$ is the period of the carrier wave, $F$ is the peak electric
field strength, $N_{\text{C}}$ is the number of cycles under
the envelope function and $k$ is a linear chirp parameter, which is calculated from the group delay dispersion ($\mathrm{GDD}$) according to the formula:
\begin{equation}
    k=\frac{\mathrm{GDD}}{\mathrm{GDD}^2+\frac{\Delta\tau^4}{8\cdot\left(\ln{2}\right)^2}},
\end{equation}
where $\Delta\tau$ is the pulse duration. In all our simulations we set $N_{\text{C}}=3$ since this corresponds to the shortest realistic pulse duration (i.e. a nearly single-cycle laser pulse, regarding intensity FWHM) and it is suitable for the generation of isolated attosecond pulses. We examine four different values of the carrier wave period: the $T=110$ corresponds to a ca. $800\;\mathrm{nm}$ near-infrared central wavelength which is very usual in attosecond physics, and $T=210$ corresponds to a ca. $1520\;\mathrm{nm}$ carrier wavelength. The $T=117$ (ca. $850\;\mathrm{nm}$) corresponds to the central wavelength of the SYLOS2 laser system, while $T=142$ (ca. $1030\;\mathrm{nm}$) is the period of the carrier wave of the HR1 laser system at ELI-ALPS \cite{ELIlaser,Ye2020beamline}.

For hydrogen, we use $Z=1$, $\mu=1$ and $\alpha^2=2$. For argon, we use $Z=1.07623$, $\mu=1$ and $\alpha^2=1.72631$ so that the ground state energy is $-0.579157$. Regarding discretization, we set typically a spatial grid size of $0.2$ and a temporal step of $\Delta t=0.01$ since these are sufficient for the numerical errors to be within line thickness. We use box boundary conditions and set the size of the box to be sufficiently large so that the reflections are kept below $10^{-8}$ atomic units.

The 3D reference results (i.e. the simulation results of the true 3D Schrödinger equation \eqref{eq:3d_tdse}) are plotted in black and are labeled ``3D reference''. The 1D simulation results and their respective colors are plotted as follows: the conventional soft-core Coulomb potential \eqref{eq:pot1_sc} in blue, the modified soft-core Coulomb potential \eqref{eq:pot1_sc_corr} in green, and the novel combined model potential \eqref{eq:pot1_gsc} in red.

\subsection{Low frequency response}

\begin{figure*}
\begin{raggedright} \hspace{4.6cm}(a)\hspace{8.7cm}(b) 

\end{raggedright}

\includegraphics[width=1\columnwidth]{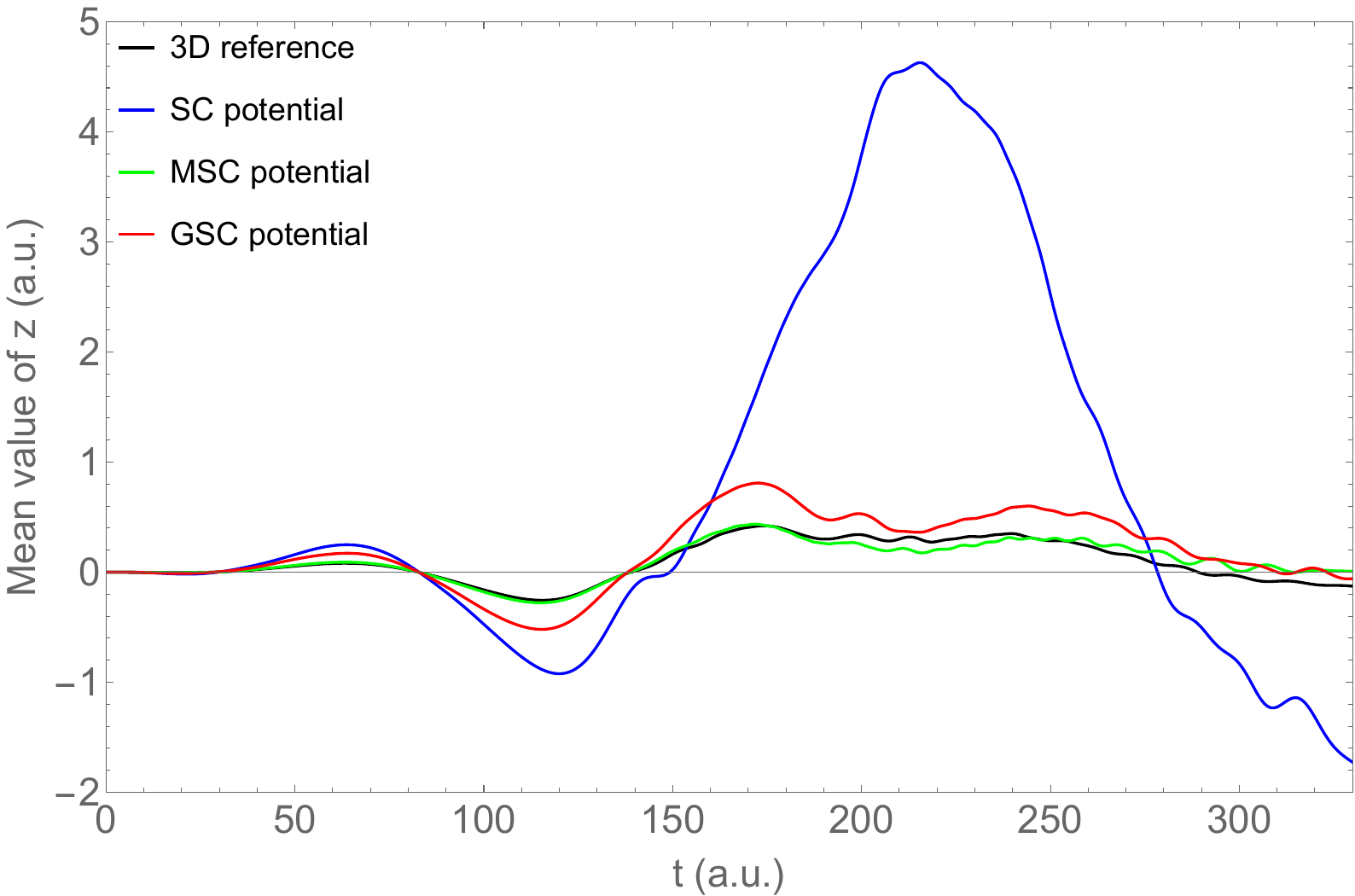}\hspace{0.5cm}\includegraphics[width=1\columnwidth]{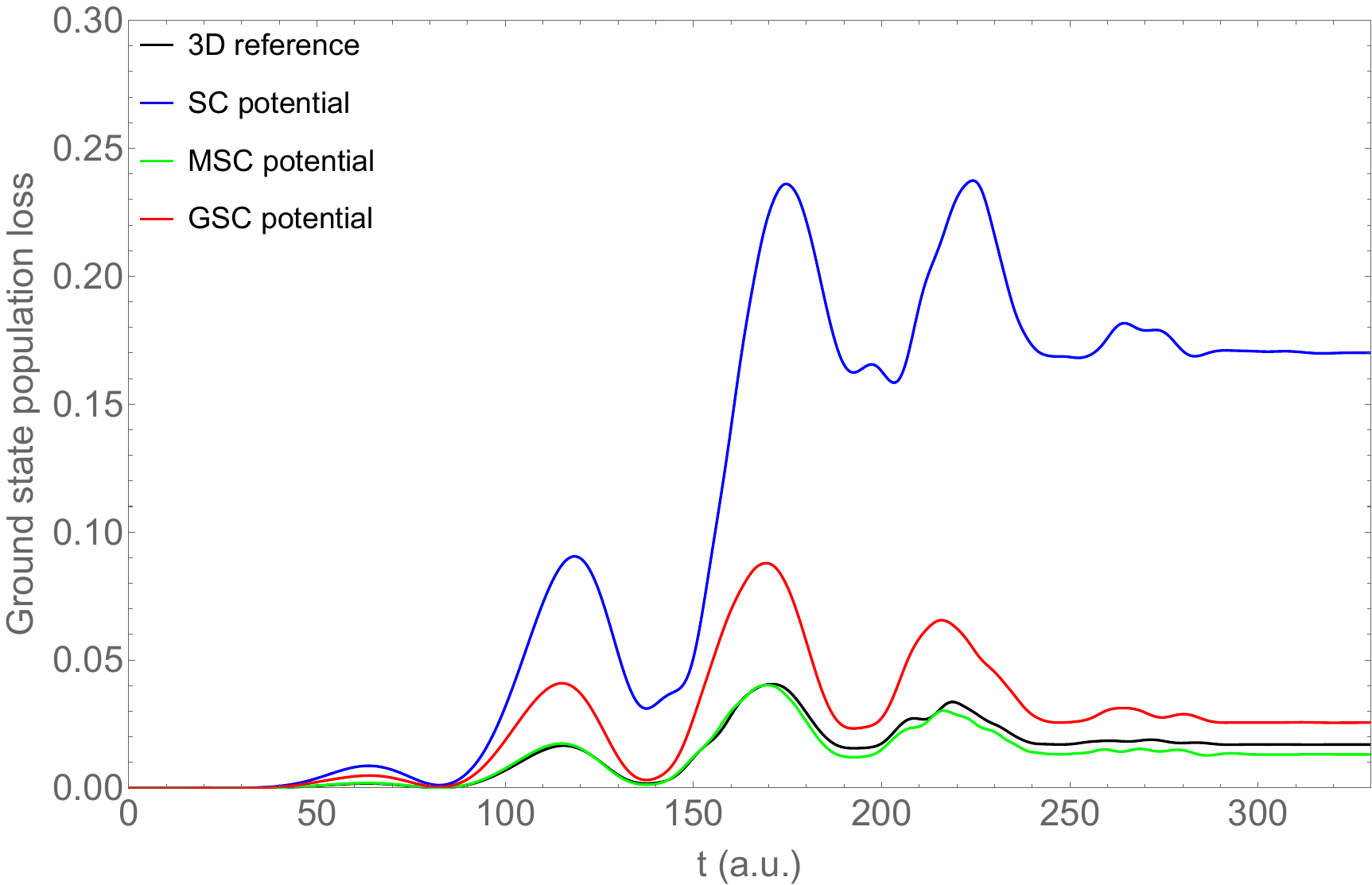}

\begin{raggedright} \hspace{4.6cm}(c)\hspace{8.7cm}(d) 

\end{raggedright}

\includegraphics[width=1\columnwidth]{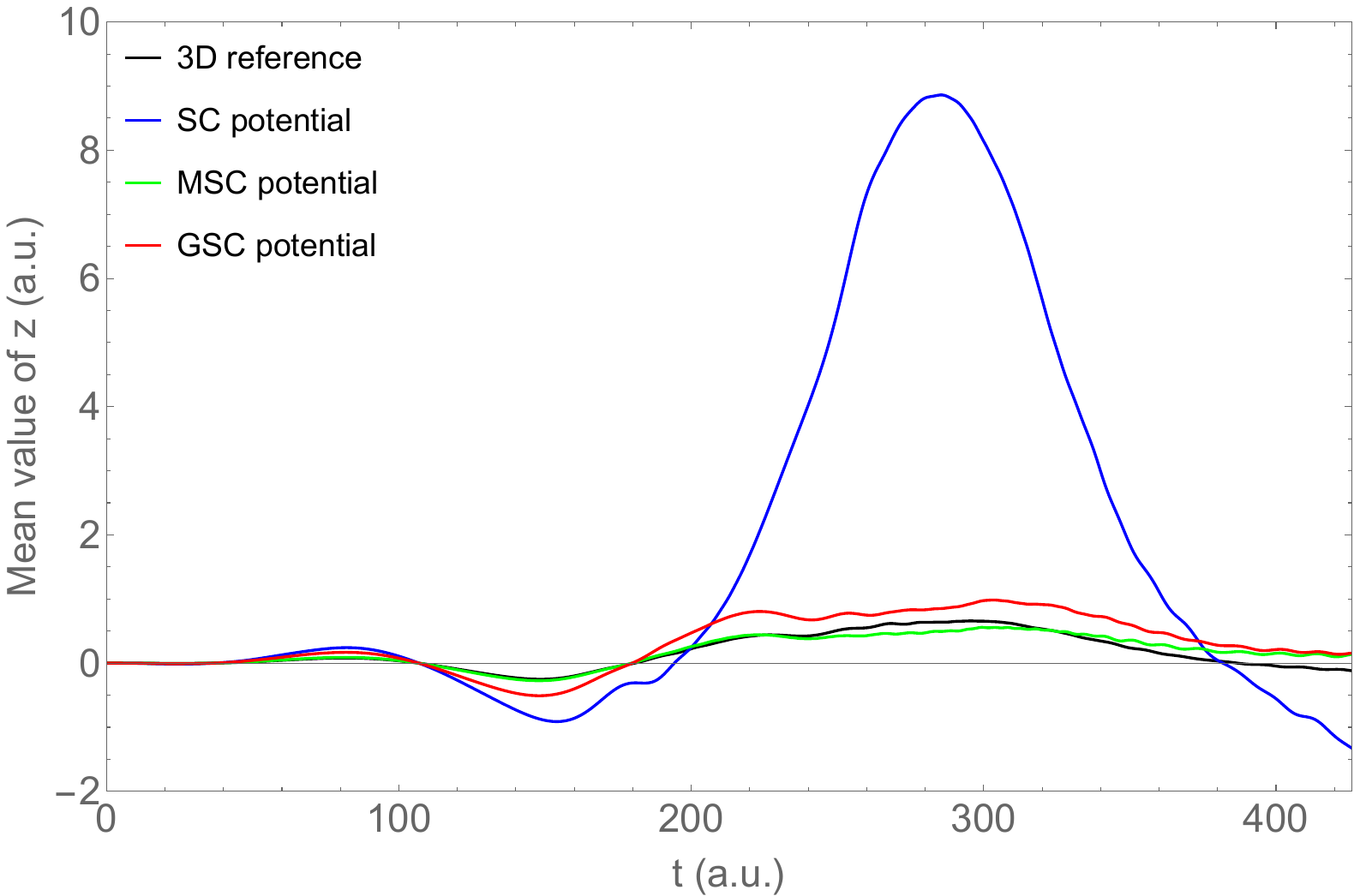}
\includegraphics[width=1\columnwidth]{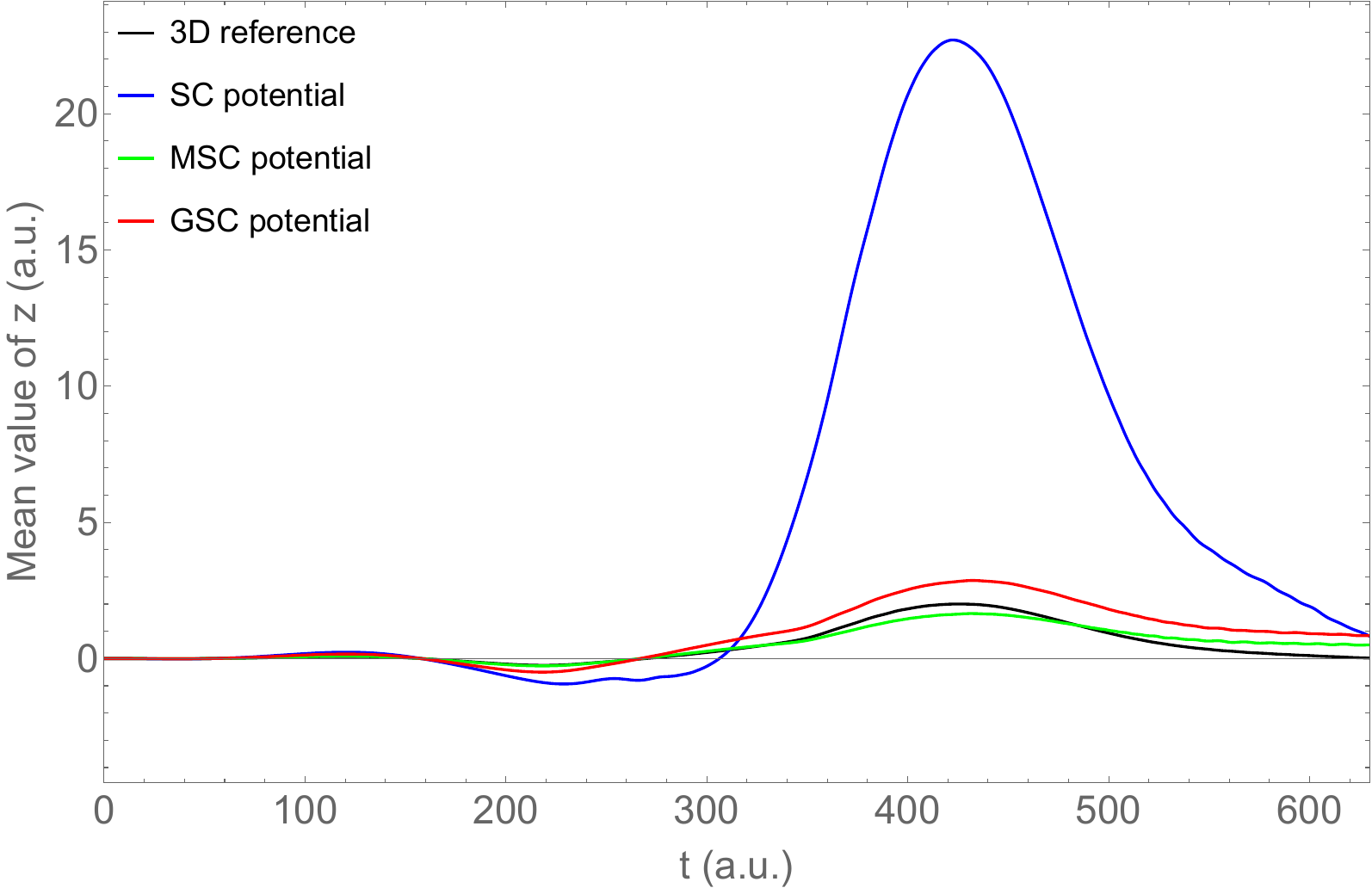}

\protect\protect\caption{Result for hydrogen, (a) time dependence of the dipole moment ($\left\langle z\right\rangle (t)$) and (b) ground state-population loss $g(t)$ with laser pulse parameters $T=110$, $N_{\text{C}}=3$, $F=0.06$. Time dependence of the dipole moment ($\left\langle z\right\rangle (t)$) with laser pulse parameters $N_{\text{C}}=3$, $F=0.06$ (c) $T=142$ and (d) $T=210$.}

\label{fig:T110_T142} 
\end{figure*}

First, we discuss the results of a laser pulse having a peak electric field value of $F=0.06$, which is in the tunnel ionization regime of hydrogen and is in the range of typical intensities used for HHG. In Fig. \ref{fig:T110_T142} we plot the corresponding time-dependent mean values $\left\langle z\right\rangle (t)$ (the magnitude of which equals the dipole moment in atomic units) for three different carrier wave period values and the ground state population loss $g(t)$, corresponding to $T=110$, for all the 1D model systems listed above. These curves show that the simulation results obtained with our GSC model potential are quantitatively closer to the 3D results than the SC potential, but a bit less precise than the MSC potential, which was devised especially for the purpose of very good accuracy for the low frequency response.

\begin{figure*}
\begin{raggedright} \hspace{4.6cm}(a)\hspace{8.7cm}(b) 

\end{raggedright}

\includegraphics[width=1\columnwidth]{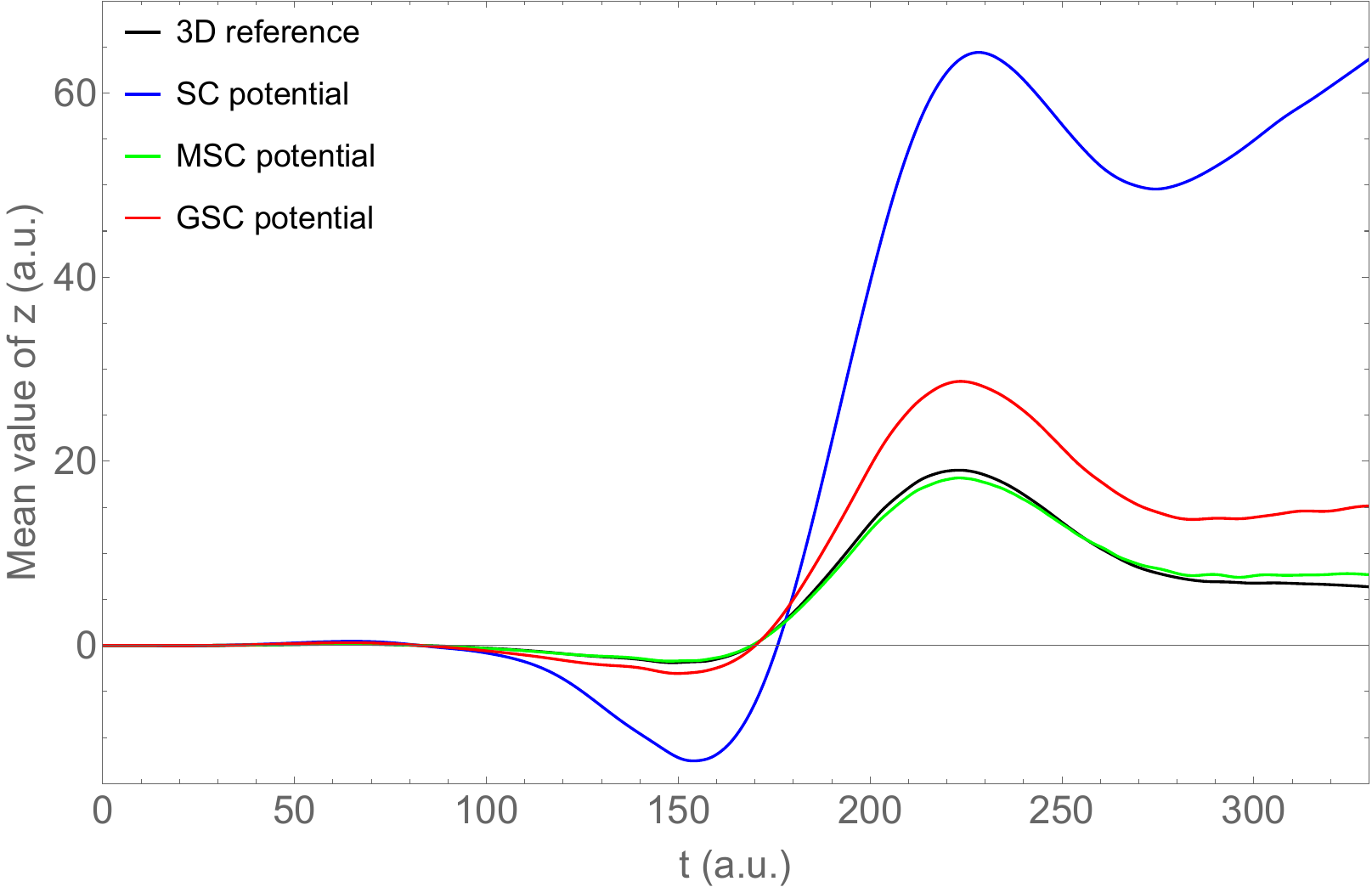}\hspace{0.5cm}\includegraphics[width=1\columnwidth]{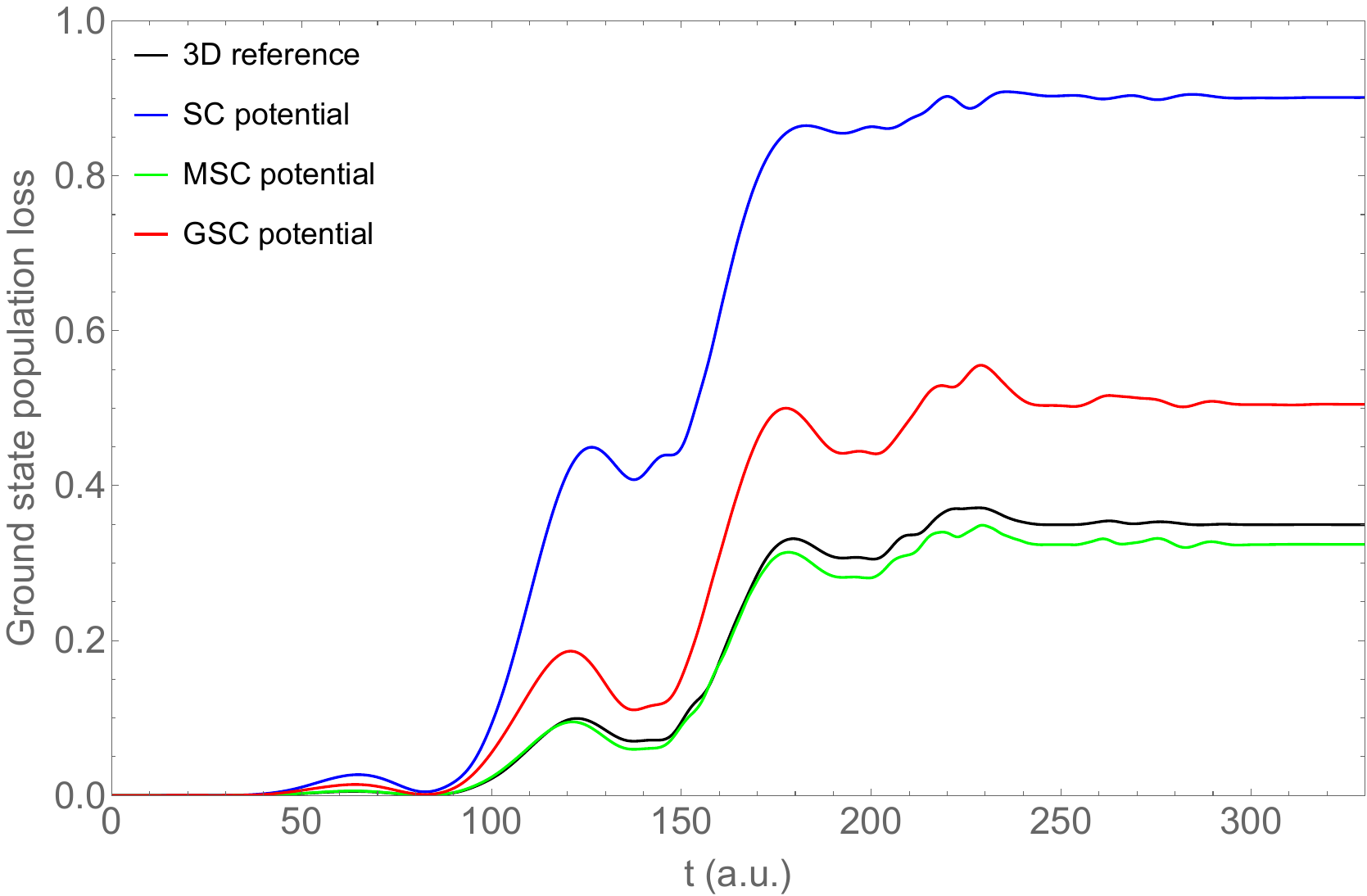}

\protect\protect\caption{Time dependence of (a) the dipole moment ($\left\langle z\right\rangle (t)$) and (b) ground-state population loss $g(t)$ using different 1D model potentials,
for hydrogen, under the influence of the same laser pulse with $F = 0.1$, $N_{\text{C}}=3$, and $T = 110$. Results of the corresponding 3D simulation are plotted in black.}

\label{fig:T110_F01} 
\end{figure*}

In order to demonstrate the capabilities of the GSC potential, in Fig. \ref{fig:T110_F01}, we also present results for $T=110$, $F=0.1$, for which the GSC potential gives good results too. However, this intensity is well above the typical tunnel ionization regime used for HHG. In accordance with this, Fig. \ref{fig:T110_F01} (b) shows that the ground state population loss is ca. a magnitude larger than in the previous case.

\begin{figure*}
\begin{raggedright} \hspace{4.6cm}(a)\hspace{8.7cm}(b) 

\end{raggedright}

\includegraphics[width=1\columnwidth]{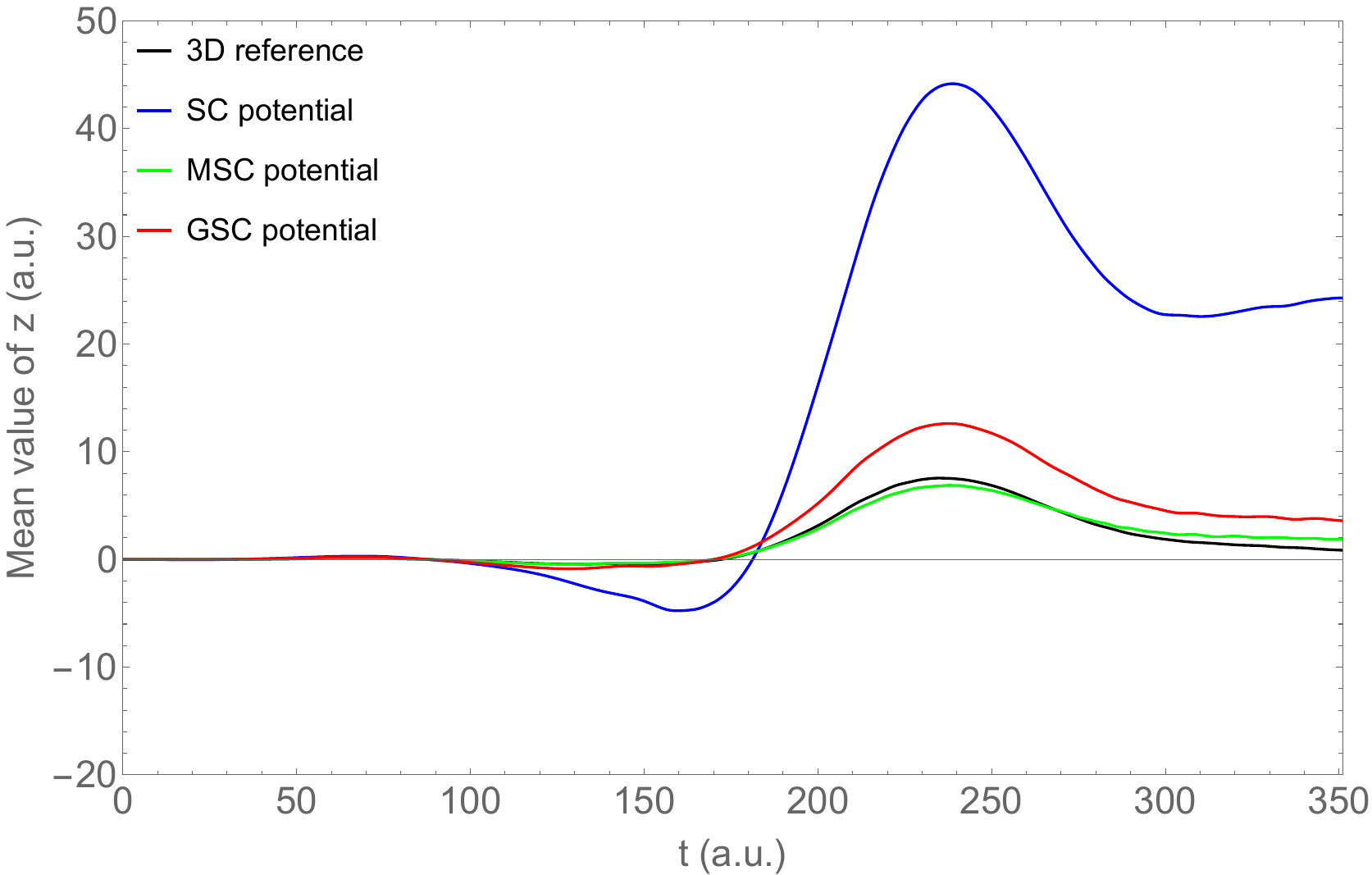}\hspace{0.5cm}\includegraphics[width=1\columnwidth]{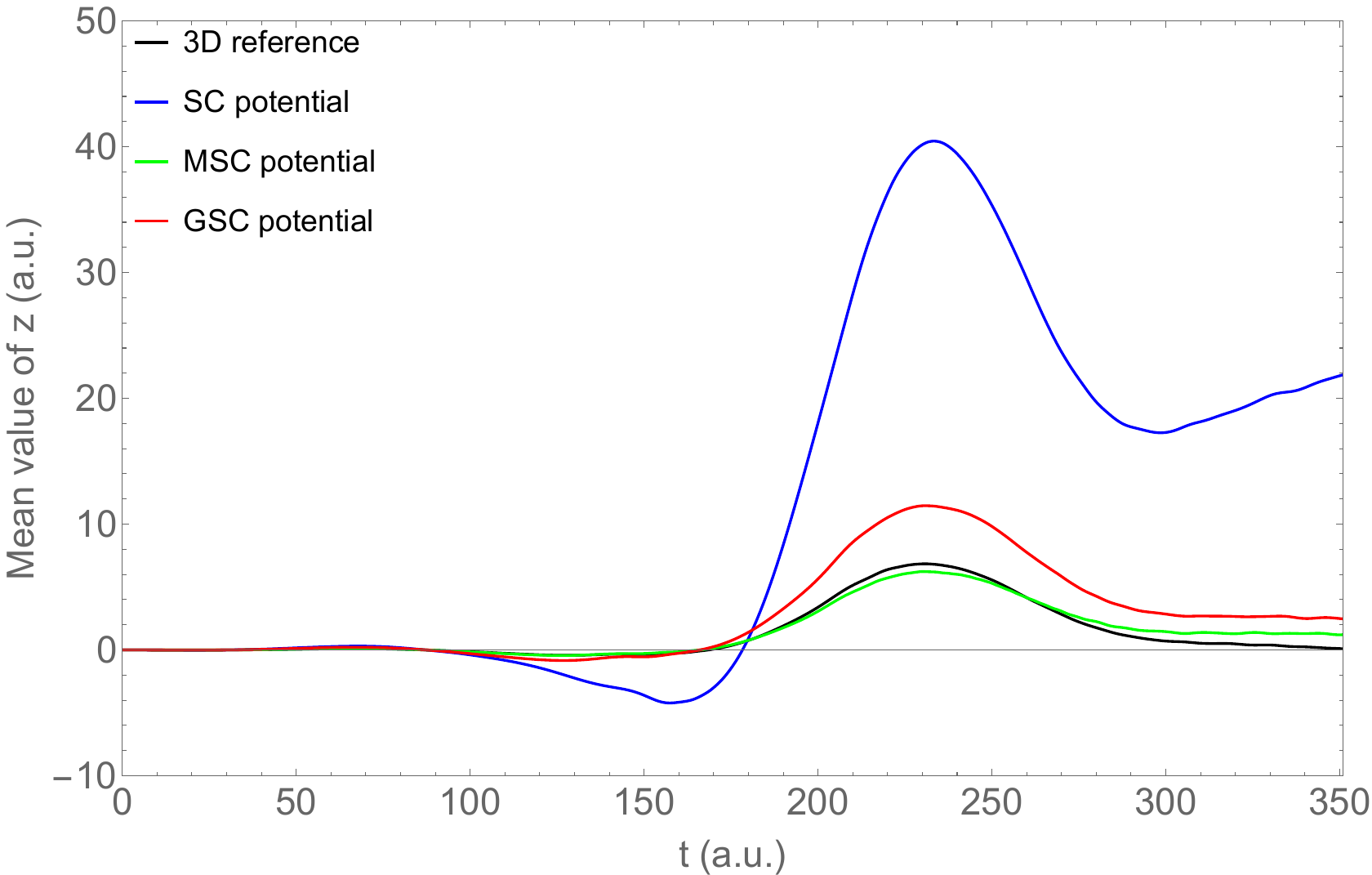}

\protect\protect\caption{Time dependence of the dipole moment ($\left\langle z\right\rangle (t)$) for argon with $T=117$, $N_{\text{C}}=3$, $F=0.1$, (a) $k=0$ and (b) $k=3.88\cdot 10^{-6}$. }

\label{fig:ar_T117} 
\end{figure*}

Fig. \ref{fig:ar_T117} shows $\left\langle z\right\rangle (t)$ for an argon atom driven by a laser pulse with parameters $T=117$, $N_{\text{C}}=3$, $F=0.1$, without chirp ($k=0$, panel (a)) and with chirp ($k=+3.8\cdot10^{-6}$, panel (b)). Here we model the 3D Ar atom in the single active electron approximation \cite{lewenstein1994hhgtheory} simply by setting the effective ion-core charge $Z=1.07623$ in order to match the ionization potential to the experimental value. For the 1D model potentials we set $\mu=1$ and $\alpha^2=1.72631$ which yields a ground state energy of $-0.579157$. The low-frequency response, and the GSC potential's accuracy, is only slightly affected by this chirp value.

\subsection{High order harmonic spectra}

The accurate computation of the high order harmonic spectrum is of fundamental importance in strong-field physics, because this represents
the highly nonlinear atomic response to the strong-field excitation,
with well-known characteristic features \cite{McPherson_JOSAB_1987_HHG,Ferray_JPhysB_1988_HHG,Harris_OptCom_1993_HHG_Atto,Krausz_RevModPhys_2009_Attosecond_physics,gombkoto2016quantoptichhg}.
Besides the high-order harmonic yield, the suitable phase relations
enable to generate attosecond pulses of XUV light \cite{Farkas_PhysLettA_1992_AttoPulse,Paul_Science_2001_Atto_pulsetrain,hentschel2001attosecond,drescher2002attosecond,kienberger2002attosecond,carrera2006attoxuv,sansone2006attosecondisolated}.

\begin{figure*}
\begin{raggedright} \hspace{4.6cm}(a)\hspace{8.7cm}(b) 

\end{raggedright}

\includegraphics[width=1\columnwidth]{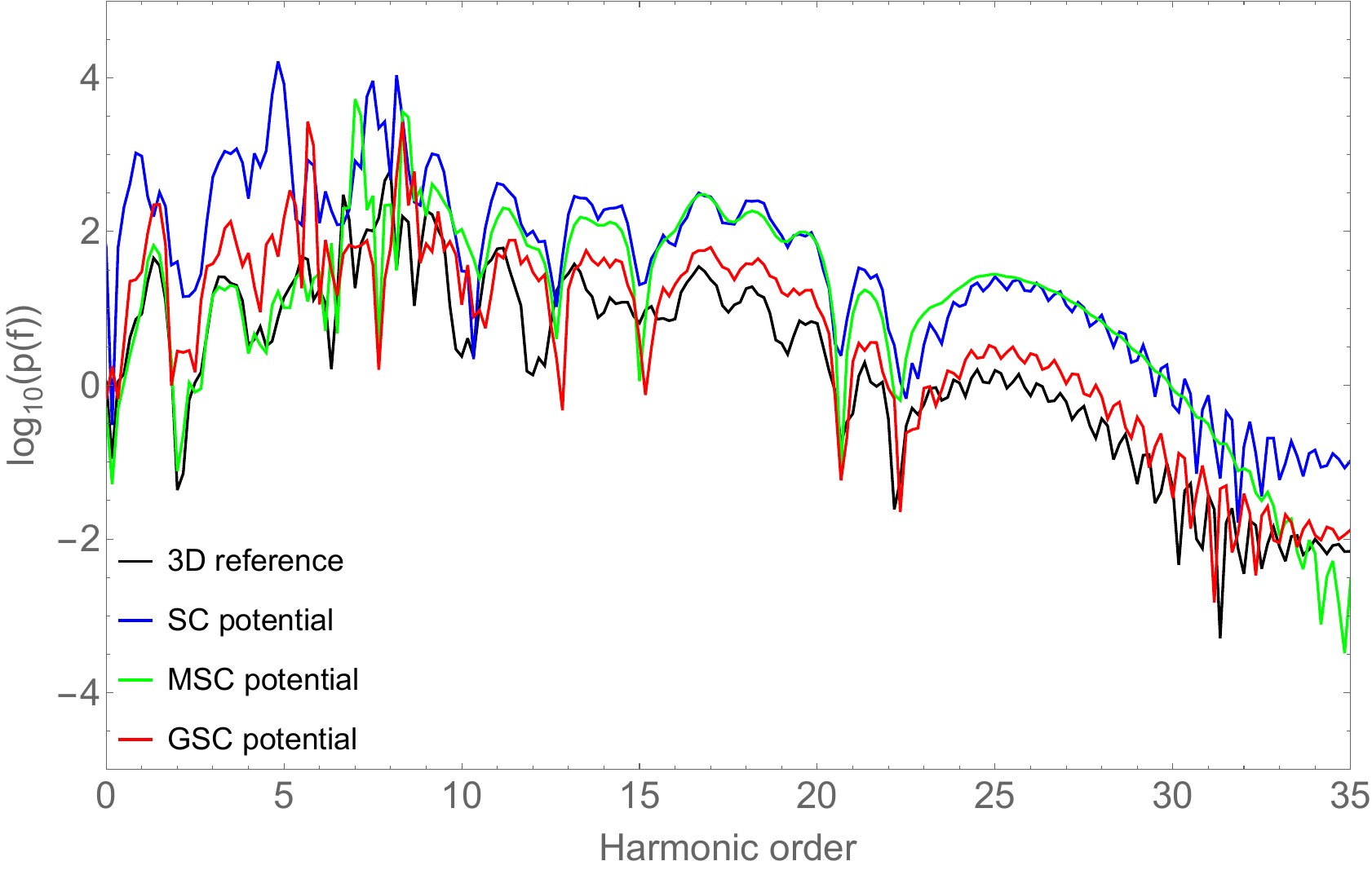}\hspace{0.5cm}\includegraphics[width=1\columnwidth]{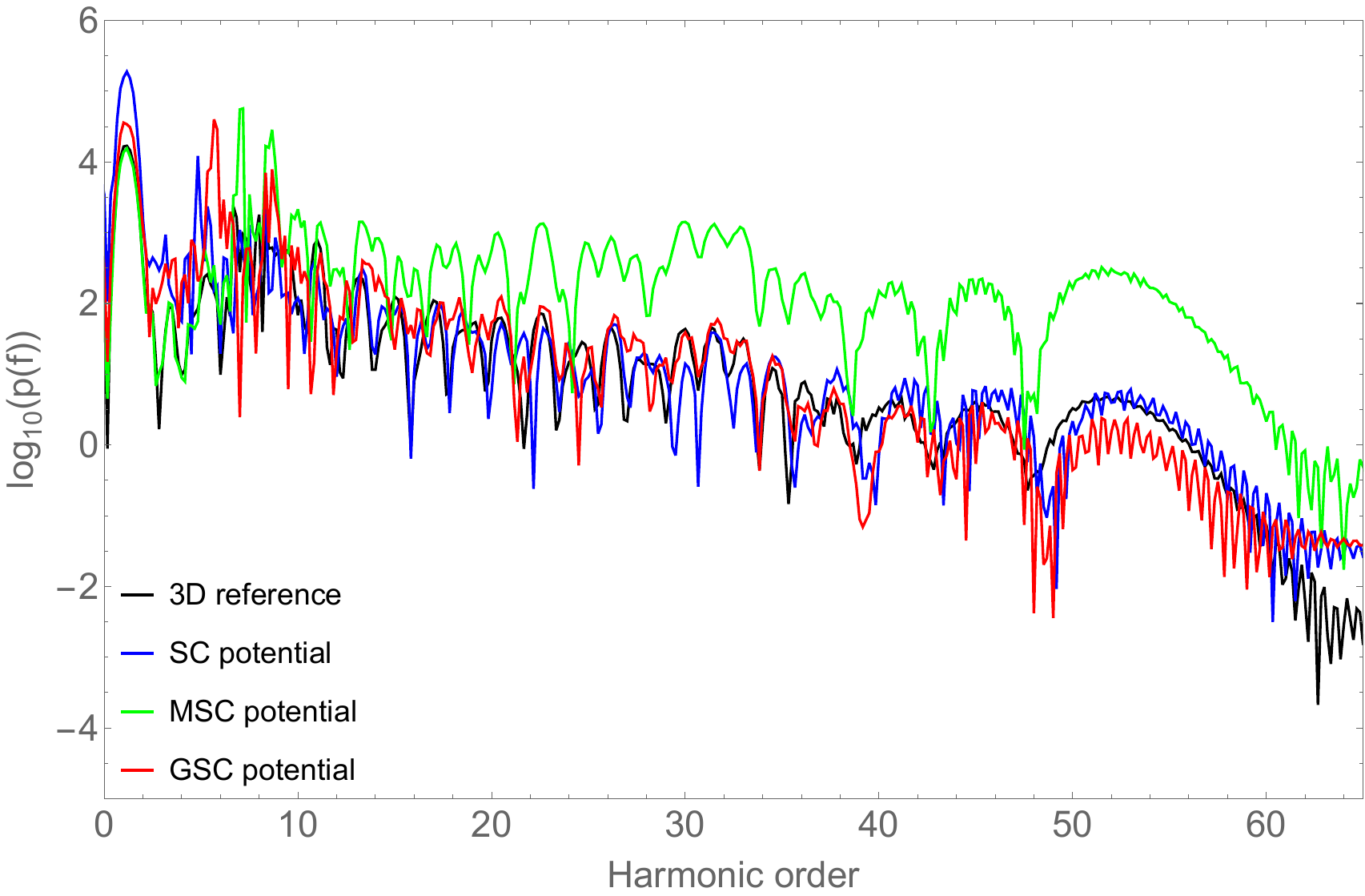}

\protect\protect\caption{The power spectrum
of the dipole acceleration for hydrogen, $T=110$, $N_{\text{C}}=3$, (a) $F=0.06$ and (b) $F=0.1$.}

\label{fig:T110_HHG} 
\end{figure*}

\begin{figure*}
\begin{raggedright} \hspace{4.6cm}(a)\hspace{8.7cm}(b) 

\end{raggedright}

\includegraphics[width=1\columnwidth]{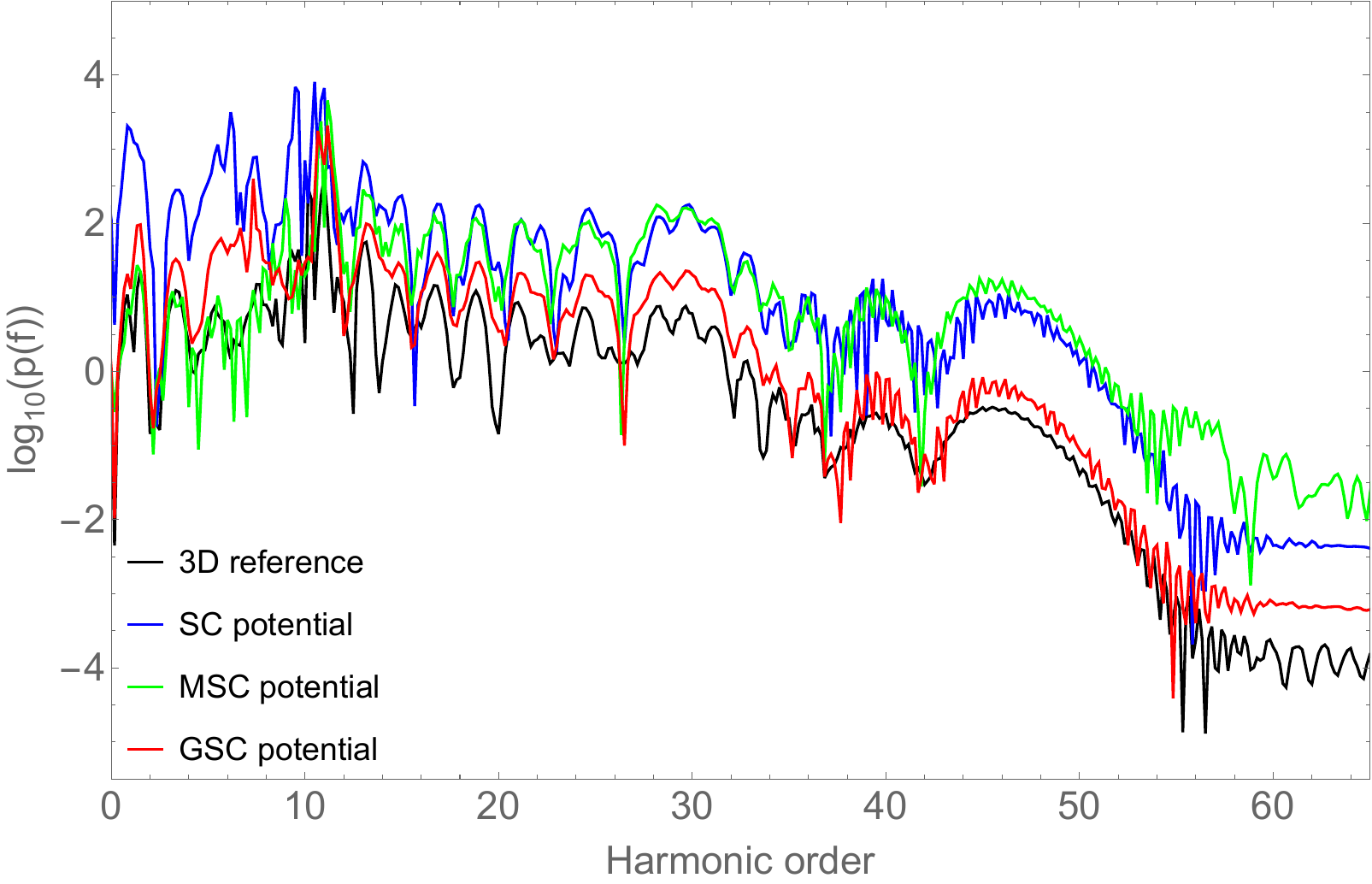}\hspace{0.5cm}\includegraphics[width=1\columnwidth]{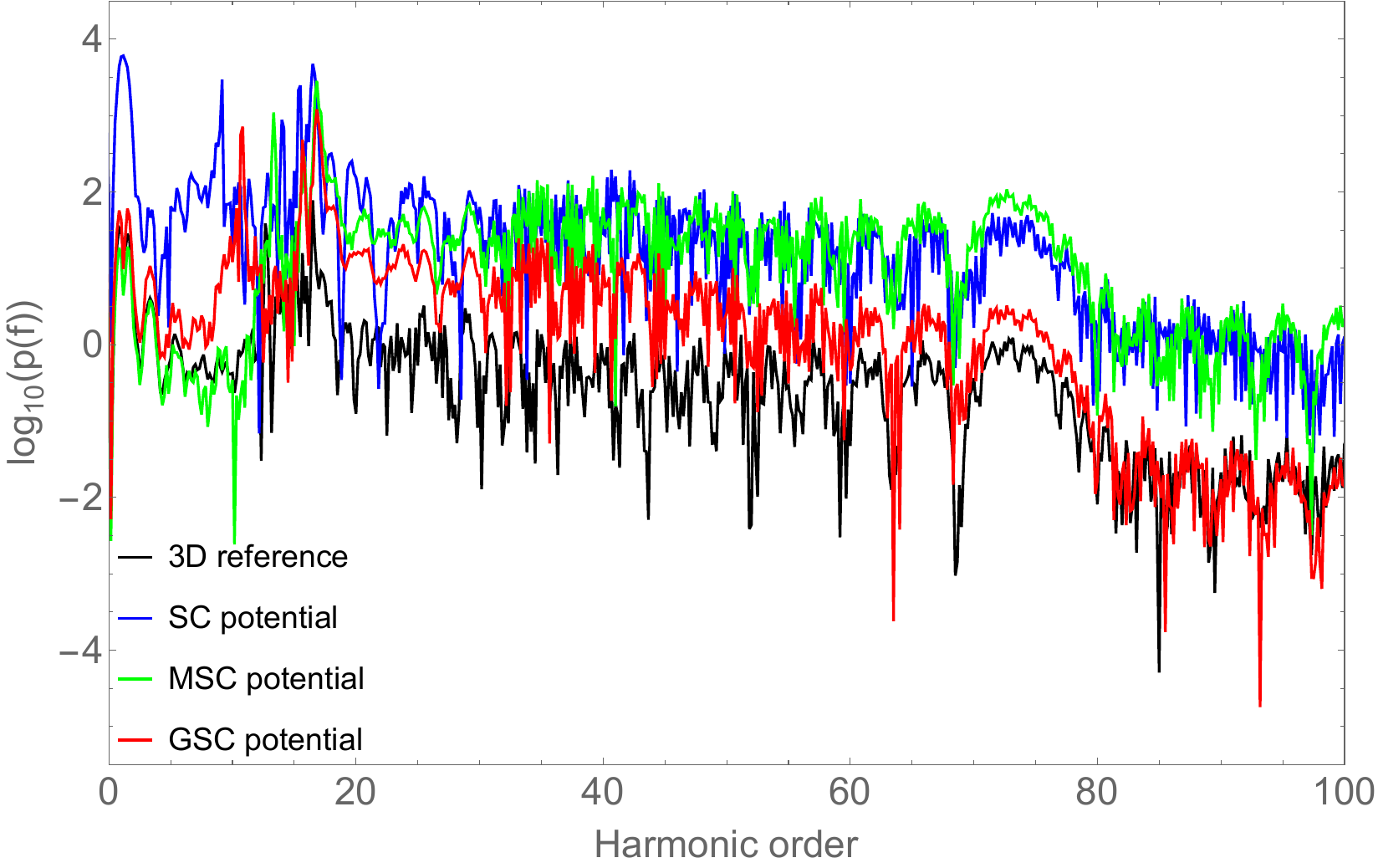}

\protect\protect\caption{The power spectrum
of the dipole acceleration for hydrogen, $N_{\text{C}}=3$, $F=0.06$, (a) $T=142$ and (b) $T=210$.}

\label{fig:T142_T210} 
\end{figure*}

\begin{figure*}
\begin{raggedright} \hspace{4.6cm}(a)\hspace{8.7cm}(b) 

\end{raggedright}

\includegraphics[width=1\columnwidth]{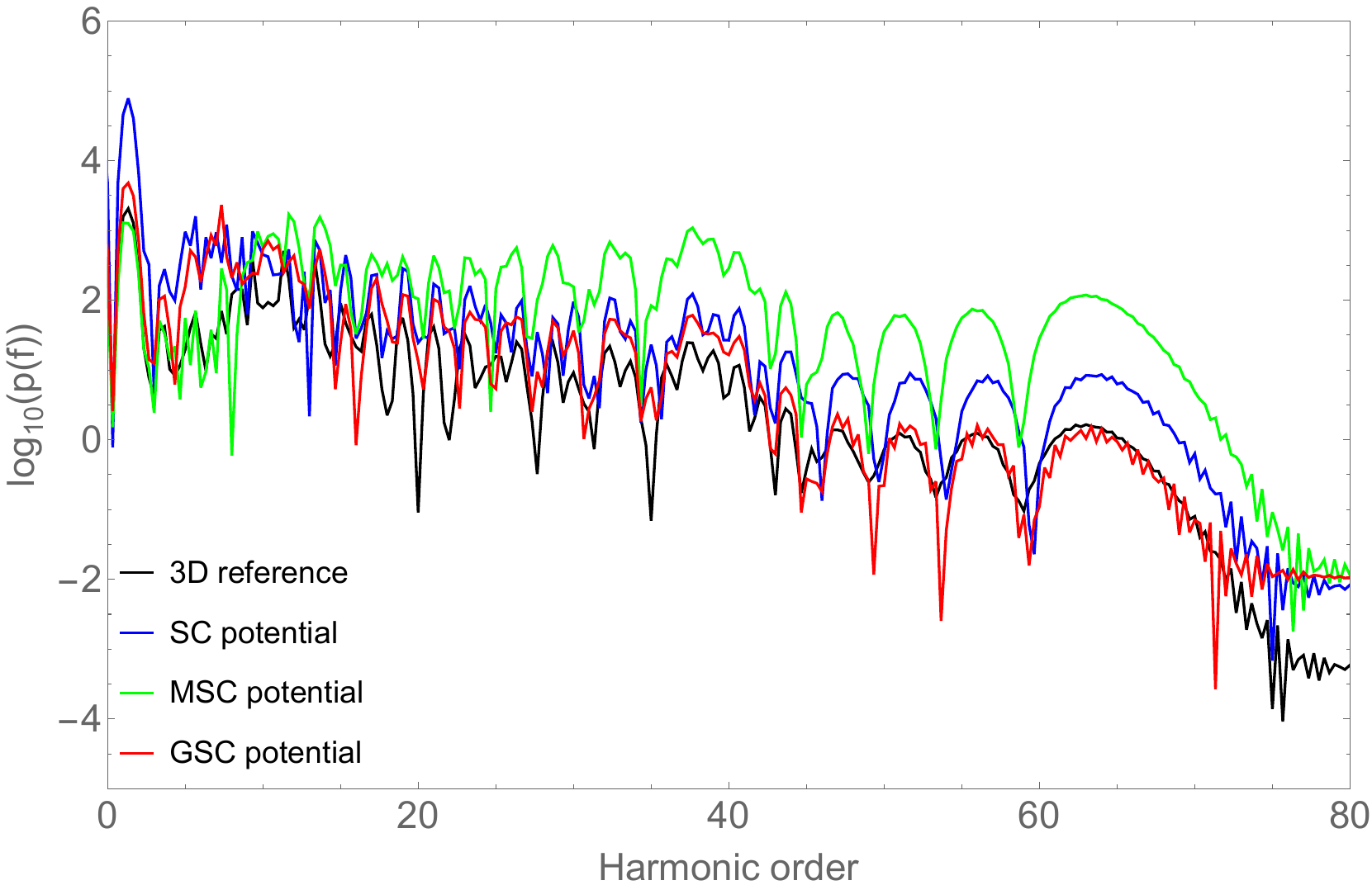}\hspace{0.5cm}\includegraphics[width=1\columnwidth]{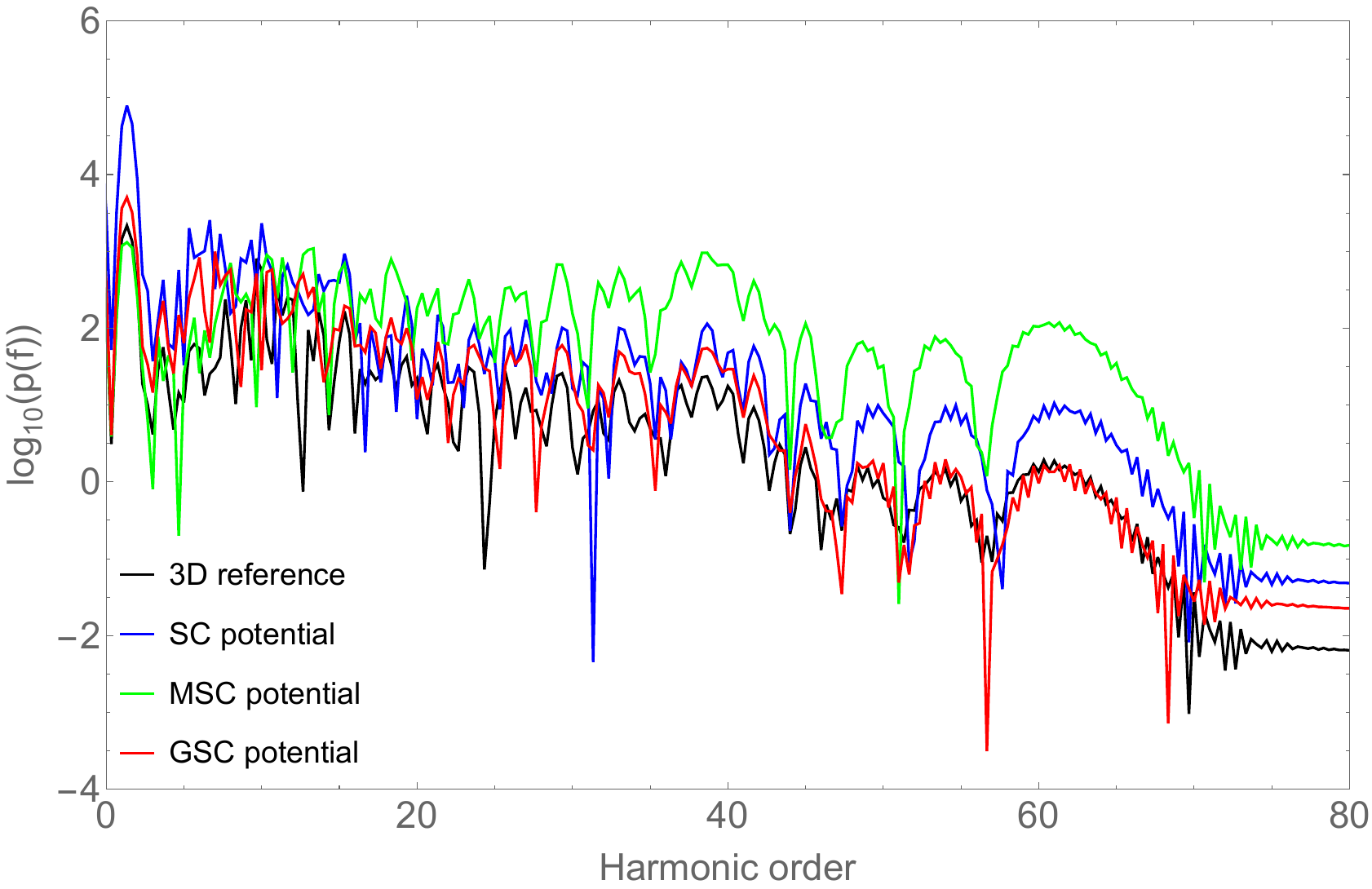}

\protect\protect\caption{The power spectrum
of the dipole acceleration for argon with $T=117$, $N_{\text{C}}=11$, $F=0.1$ with (a) $k=0$ and (b) $k=3.88\cdot 10^{-6}$.}

\label{fig:ar_T117_HHG} 
\end{figure*}

In Fig. \ref{fig:T110_HHG} (a), we plot the power spectrum
of the dipole acceleration (see Eq. \ref{eq:z_spectrumpow}) for hydrogen with the
parameters corresponding to Fig. \ref{fig:T110_T142} (a) and (b), while Fig. \ref{fig:T110_HHG} (b) corresponds to Fig. \ref{fig:T110_F01}. These results show that the GSC potential has considerably better overall modelling capability of the HHG spectra than the SC or MSC potentials.
Keeping the lower field strength of $F=0.06$ and $N_{\text{C}}=3$, we compare the HHG spectra corresponding to $T=142$ and $T=210$ in Fig. \ref{fig:T142_T210}. 
For $T=142$ (corresponding to 1030 nm carrier wavelength), the quality of the results is still good, but the case of  $T=210$ (i.e. 1520 nm) is beyond reach also for the GSC potential, except for the cutoff region.
Regarding Ar, 
we plot the HHG spectra with laser pulse parameters $T=117$, $N_{\text{C}}=3$, $F=0.01$, $k=0$ and $k=3.88\cdot 10^{-6}$ in Fig. \ref{fig:ar_T117_HHG}. 
For these cases, the results obtained with the GSC potential agree very well with the 3D reference simulation result for all the harmonics.

\section{Discussion and conclusions}

The results presented in the previous section demonstrate that it
is possible to quantitatively model the high order harmonic spectrum with suitable 1D model potentials if the laser polarization is linear.
The overall best results are obtained with the novel GSC model potential \eqref{eq:pot1_gsc} which is also very effective to use numerically. This means that we can perform quantum simulations of a single active electron atom driven by a strong linearly polarized laser pulse during a couple of minutes and obtain a fairly accurate HHG spectrum.

We expect that this novel potential can also be successfully used as a building block in the 1D quantum modelling and simulation  of somewhat larger atomic systems, and it may also lead to more effective ensemble simulations of HHG spectra based on quantum dynamics.

\begin{acknowledgments}
Krisztina Sallai was supported by the UNKP-23-3 New National Excellence Program of the Ministry of Human Capacities of Hungary. 
The ELI-ALPS project (GINOP-2.3.6-15-2015-00001) is supported by the European Union and co-financed by the European Regional Development Fund.
\end{acknowledgments}

\appendix

\section{
\label{sec:comparable-physical-quantities}}

For completeness, we list here the physical quantities that we use
for characterizing the strong-field process, both in 1D and 3D.

\begin{equation}
\varrho_{z}^{{\rm 1D}}(z,t)=\left|\Psi^{{\rm 1D}}(z,t)\right|^{2}.\label{eq:z_density1}
\end{equation}
We calculate the mean value of $z$ as 
\begin{equation}
\left\langle z\right\rangle (t)=\int_{-\infty}^{\infty}z\varrho_{z}(z,t)\text{d}z,\label{eq:z_mean_z}
\end{equation}
the mean value of the $z$-velocity and the $z$-acceleration using
the Ehrenfest theorems as 
\begin{equation}
\left\langle v_{z}\right\rangle (t)=\frac{\partial\left\langle z\right\rangle }{\partial t},\,\,\,\,\,\,\,\left\langle a_{z}\right\rangle (t)=\frac{\partial\left\langle v_{z}\right\rangle }{\partial t},\label{eq:z_mean_vz}
\end{equation}
in both the 3D and the 1D cases. It is also interesting to determine
the ground state population loss 
\begin{equation}
g(t)=1-\left|\left\langle \Psi(0)|\Psi(t)\right\rangle \right|^{2},\label{eq:proj_ground_loss}
\end{equation}
even though this refers to the population losses of two completely
different states in 1D and 3D. 
\newline We calculate the spectrum from the dipole acceleration $\left\langle a_{z}\right\rangle $,
and then the power spectrum as 
\begin{equation}
p(f)=\left|\mathcal{F}\left[\left\langle a_{z}\right\rangle \right](f)\right|^{2}\label{eq:z_spectrumpow}
\end{equation}
where $\mathcal{F}$ denotes the Fourier transform and $f$ is its
frequency variable.

\bibliographystyle{unsrtnat}
\bibliography{0Bibliography,2Bibliography,3Bibliography,bib}

\begin{thebibliography}{62}
\providecommand{\natexlab}[1]{#1}
\providecommand{\url}[1]{\texttt{#1}}
\expandafter\ifx\csname urlstyle\endcsname\relax
  \providecommand{\doi}[1]{doi: #1}\else
  \providecommand{\doi}{doi: \begingroup \urlstyle{rm}\Url}\fi

\bibitem[Hentschel et~al.(2001)Hentschel, Kienberger, Spielmann, Reider,
  Milosevic, Brabec, Corkum, Heinzmann, Drescher, and
  Krausz]{hentschel2001attosecond}
M~Hentschel, R~Kienberger, Ch~Spielmann, Georg~A Reider, N~Milosevic, Thomas
  Brabec, Paul Corkum, Ulrich Heinzmann, Markus Drescher, and Ferenc Krausz.
\newblock Attosecond metrology.
\newblock \emph{Nature}, 414\penalty0 (6863):\penalty0 509--513, 2001.

\bibitem[Kienberger et~al.(2002)Kienberger, Hentschel, Uiberacker, Spielmann,
  Kitzler, Scrinzi, Wieland, Westerwalbesloh, Kleineberg, Heinzmann,
  et~al.]{kienberger2002attosecond}
Reinhard Kienberger, Michael Hentschel, Matthias Uiberacker, Ch~Spielmann,
  Markus Kitzler, Armin Scrinzi, M~Wieland, Th~Westerwalbesloh, U~Kleineberg,
  Ulrich Heinzmann, et~al.
\newblock Steering attosecond electron wave packets with light.
\newblock \emph{Science}, 297\penalty0 (5584):\penalty0 1144--1148, 2002.

\bibitem[Drescher et~al.(2002)Drescher, Hentschel, Kienberger, Uiberacker,
  Yakovlev, Scrinzi, Westerwalbesloh, Kleineberg, Heinzmann, and
  Krausz]{drescher2002attosecond}
Markus Drescher, Michael Hentschel, R~Kienberger, Matthias Uiberacker,
  Vladislav Yakovlev, Armin Scrinzi, Th~Westerwalbesloh, U~Kleineberg, Ulrich
  Heinzmann, and Ferenc Krausz.
\newblock Time-resolved atomic inner-shell spectroscopy.
\newblock \emph{Nature}, 419\penalty0 (6909):\penalty0 803--807, 2002.

\bibitem[Baltu{\v{s}}ka et~al.(2003)Baltu{\v{s}}ka, Udem, Uiberacker,
  Hentschel, Goulielmakis, Gohle, Holzwarth, Yakovlev, Scrinzi, H{\"a}nsch,
  et~al.]{baltuska2003attosecond}
Andrius Baltu{\v{s}}ka, Th~Udem, M~Uiberacker, M~Hentschel, E~Goulielmakis,
  Ch~Gohle, Ronald Holzwarth, VS~Yakovlev, A~Scrinzi, TW~H{\"a}nsch, et~al.
\newblock Attosecond control of electronic processes by intense light fields.
\newblock \emph{Nature}, 421\penalty0 (6923):\penalty0 611--615, 2003.

\bibitem[Uiberacker et~al.(2007)Uiberacker, Uphues, Schultze, Verhoef,
  Yakovlev, Kling, Rauschenberger, Kabachnik, Schr{\"o}der, Lezius,
  et~al.]{Uiberacker_Nature_2007}
Matthias Uiberacker, Th~Uphues, Martin Schultze, Aart~Johannes Verhoef,
  Vladislav Yakovlev, Matthias~F Kling, Jens Rauschenberger, Nicolai~M
  Kabachnik, Hartmut Schr{\"o}der, Matthias Lezius, et~al.
\newblock Attosecond real-time observation of electron tunnelling in atoms.
\newblock \emph{Nature}, 446\penalty0 (7136):\penalty0 627--632, 2007.

\bibitem[Krausz and Ivanov(2009)]{Krausz_RevModPhys_2009_Attosecond_physics}
Ferenc Krausz and Misha Ivanov.
\newblock Attosecond physics.
\newblock \emph{Reviews of Modern Physics}, 81\penalty0 (1):\penalty0 163--234,
  2009.

\bibitem[Hommelhoff et~al.(2009)Hommelhoff, Kealhofer, Aghajani-Talesh,
  Sortais, Foreman, and Kasevich]{hommelhoff2009extremelocalization}
Peter Hommelhoff, Catherine Kealhofer, Anoush Aghajani-Talesh, Yvan~RP Sortais,
  Seth~M Foreman, and Mark~A Kasevich.
\newblock Extreme localization of electrons in space and time.
\newblock \emph{Ultramicroscopy}, 109\penalty0 (5):\penalty0 423--429, 2009.

\bibitem[Schultze et~al.(2010)Schultze, Fie{\ss}, Karpowicz, Gagnon, Korbman,
  Hofstetter, Neppl, Cavalieri, Komninos, Mercouris,
  et~al.]{Schultze_Science_2010}
Martin Schultze, Markus Fie{\ss}, Nicholas Karpowicz, Justin Gagnon, Michael
  Korbman, Michael Hofstetter, S~Neppl, Adrian~L Cavalieri, Yannis Komninos,
  Th~Mercouris, et~al.
\newblock Delay in photoemission.
\newblock \emph{Science}, 328\penalty0 (5986):\penalty0 1658--1662, 2010.

\bibitem[Haessler et~al.(2010)Haessler, Caillat, Boutu, Giovanetti-Teixeira,
  Ruchon, Auguste, Diveki, Breger, Maquet, Carr{\'e},
  et~al.]{Haessler_NatPhys_2010}
Stefan Haessler, J~Caillat, W~Boutu, C~Giovanetti-Teixeira, T~Ruchon,
  T~Auguste, Z~Diveki, P~Breger, A~Maquet, B~Carr{\'e}, et~al.
\newblock Attosecond imaging of molecular electronic wavepackets.
\newblock \emph{Nature Physics}, 6\penalty0 (3):\penalty0 200--206, 2010.

\bibitem[Pfeiffer et~al.(2012)Pfeiffer, Cirelli, Smolarski, Dimitrovski,
  Abu-Samha, Madsen, and Keller]{pfeiffer2012attoclock}
Adrian~N Pfeiffer, Claudio Cirelli, Mathias Smolarski, Darko Dimitrovski,
  Mahmoud Abu-Samha, Lars~Bojer Madsen, and Ursula Keller.
\newblock Attoclock reveals natural coordinates of the laser-induced tunnelling
  current flow in atoms.
\newblock \emph{Nature Physics}, 8\penalty0 (1):\penalty0 76--80, 2012.

\bibitem[Shafir et~al.(2012)Shafir, Soifer, Bruner, Dagan, Mairesse,
  Patchkovskii, Ivanov, Smirnova, and Dudovich]{shafir2012tunneling}
Dror Shafir, Hadas Soifer, Barry~D Bruner, Michal Dagan, Yann Mairesse, Serguei
  Patchkovskii, Misha~Yu Ivanov, Olga Smirnova, and Nirit Dudovich.
\newblock Resolving the time when an electron exits a tunnelling barrier.
\newblock \emph{Nature}, 485\penalty0 (7398):\penalty0 343--346, 2012.

\bibitem[Ranitovic et~al.(2014)Ranitovic, Hogle, Rivi{\`e}re, Palacios, Tong,
  Toshima, Gonz{\'a}lez-Castrillo, Martin, Mart{\'\i}n, Murnane,
  et~al.]{ranitovic2014attosecondcontrol}
Predrag Ranitovic, Craig~W Hogle, Paula Rivi{\`e}re, Alicia Palacios, Xiao-Ming
  Tong, Nobuyuki Toshima, Alberto Gonz{\'a}lez-Castrillo, Leigh Martin,
  Fernando Mart{\'\i}n, Margaret~M Murnane, et~al.
\newblock Attosecond vacuum uv coherent control of molecular dynamics.
\newblock \emph{Proceedings of the National Academy of Sciences}, 111\penalty0
  (3):\penalty0 912--917, 2014.

\bibitem[Peng et~al.(2015)Peng, Jiang, Geng, Xiong, and
  Gong]{peng2015attosecondtracing}
Liang-You Peng, Wei-Chao Jiang, Ji-Wei Geng, Wei-Hao Xiong, and Qihuang Gong.
\newblock Tracing and controlling electronic dynamics in atoms and molecules by
  attosecond pulses.
\newblock \emph{Physics Reports}, 575:\penalty0 1--71, 2015.

\bibitem[Ciappina et~al.(2017)Ciappina, P{\'e}rez-Hern{\'a}ndez, Landsman,
  Okell, Zherebtsov, F{\"o}rg, Sch{\"o}tz, Seiffert, Fennel, Shaaran,
  et~al.]{ciappina2017attosecondnano}
Marcello~F Ciappina, JA~P{\'e}rez-Hern{\'a}ndez, AS~Landsman, WA~Okell, Sergey
  Zherebtsov, Benjamin F{\"o}rg, Johannes Sch{\"o}tz, L~Seiffert, T~Fennel,
  T~Shaaran, et~al.
\newblock Attosecond physics at the nanoscale.
\newblock \emph{Reports on Progress in Physics}, 80\penalty0 (5):\penalty0
  054401, 2017.

\bibitem[Keldysh(1965)]{Keldysh_JETP_1965}
LV~Keldysh.
\newblock Ionization in the field of a strong electromagnetic wave.
\newblock \emph{Soviet Physics JETP}, 20\penalty0 (5):\penalty0 1307--1314,
  1965.

\bibitem[Varr{\'o} and Ehlotzky(1993)]{varro1993multiphotonequation}
S{\'a}ndor Varr{\'o} and F~Ehlotzky.
\newblock A new integral equation for treating high-intensity multiphoton
  processes.
\newblock \emph{Il Nuovo Cimento D}, 15\penalty0 (11):\penalty0 1371--1396,
  1993.

\bibitem[Lewenstein et~al.(1994)Lewenstein, Balcou, Ivanov, {L'Huillier}, and
  Corkum]{lewenstein1994hhgtheory}
Maciej Lewenstein, Ph~Balcou, M~Yu Ivanov, A~{L'Huillier}, and Paul~B Corkum.
\newblock Theory of high-harmonic generation by low-frequency laser fields.
\newblock \emph{Physical Review A}, 49\penalty0 (3):\penalty0 2117, 1994.

\bibitem[Protopapas et~al.(1997)Protopapas, Lappas, and
  Knight]{protopapas1997tdseionization}
M~Protopapas, DG~Lappas, and PL~Knight.
\newblock Strong field ionization in arbitrary laser polarizations.
\newblock \emph{Physical Review Letters}, 79\penalty0 (23):\penalty0 4550,
  1997.

\bibitem[Ivanov et~al.(2005)Ivanov, Spanner, and
  Smirnova]{ivanov2005strongfield}
Misha~Yu Ivanov, Michael Spanner, and Olga Smirnova.
\newblock Anatomy of strong field ionization.
\newblock \emph{Journal of Modern Optics}, 52\penalty0 (2-3):\penalty0
  165--184, 2005.

\bibitem[Gordon et~al.(2005)Gordon, Santra, and
  K{\"a}rtner]{gordon2005tdsecoulomb}
Ariel Gordon, Robin Santra, and Franz~X K{\"a}rtner.
\newblock Role of the {Coulomb} singularity in high-order harmonic generation.
\newblock \emph{Physical Review A}, 72\penalty0 (6):\penalty0 063411, 2005.

\bibitem[Frolov et~al.(2012)Frolov, Manakov, Popov, Tikhonova, Volkova, Silaev,
  Vvedenskii, and Starace]{frolov2012attosecondanalytic}
MV~Frolov, NL~Manakov, AM~Popov, OV~Tikhonova, EA~Volkova, AA~Silaev,
  NV~Vvedenskii, and Anthony~F Starace.
\newblock Analytic theory of high-order-harmonic generation by an intense
  few-cycle laser pulse.
\newblock \emph{Physical Review A}, 85\penalty0 (3):\penalty0 033416, 2012.

\bibitem[Javanainen et~al.(1988)Javanainen, Eberly, and
  Su]{eberly1988softcoulombspecra}
Juha Javanainen, Joseph~H Eberly, and Qichang Su.
\newblock Numerical simulations of multiphoton ionization and above-threshold
  electron spectra.
\newblock \emph{Physical Review A}, 38\penalty0 (7):\penalty0 3430, 1988.

\bibitem[Su and Eberly(1991)]{eberly1991softcoulombatom}
Q~Su and JH~Eberly.
\newblock Model atom for multiphoton physics.
\newblock \emph{Physical Review A}, 44\penalty0 (9):\penalty0 5997, 1991.

\bibitem[Bauer(1997)]{bauer1997tdse1dhemodel}
D~Bauer.
\newblock Two-dimensional, two-electron model atom in a laser pulse: Exact
  treatment, single-active-electron analysis, time-dependent density-functional
  theory, classical calculations, and nonsequential ionization.
\newblock \emph{Physical Review A}, 56\penalty0 (4):\penalty0 3028, 1997.

\bibitem[Chiril{\u{a}} et~al.(2010)Chiril{\u{a}}, Dreissigacker, van~der Zwan,
  and Lein]{chirilua2010HHGemission}
CC~Chiril{\u{a}}, Ingo Dreissigacker, Elmar~V van~der Zwan, and Manfred Lein.
\newblock Emission times in high-order harmonic generation.
\newblock \emph{Physical Review A}, 81\penalty0 (3):\penalty0 033412, 2010.

\bibitem[Silaev et~al.(2010)Silaev, Ryabikin, and
  Vvedenskii]{silaev2010tdsecoulombs}
AA~Silaev, M~Yu Ryabikin, and NV~Vvedenskii.
\newblock Strong-field phenomena caused by ultrashort laser pulses: Effective
  one-and two-dimensional quantum-mechanical descriptions.
\newblock \emph{Physical Review A}, 82\penalty0 (3):\penalty0 033416, 2010.

\bibitem[Sveshnikov and Khomovskii(2012)]{sveshnikov2012schrodingercoulomb}
Konstantin~Alekseevich Sveshnikov and Dmitrii~Igorevich Khomovskii.
\newblock {Schr{\"o}dinger} and {Dirac} particles in quasi-one-dimensional
  systems with a {Coulomb} interaction.
\newblock \emph{Theoretical and Mathematical Physics}, 173\penalty0
  (2):\penalty0 1587--1603, 2012.

\bibitem[Gr{\"a}fe et~al.(2012)Gr{\"a}fe, Doose, and
  Burgd{\"o}rfer]{graefe2012quantumphasespace}
Stefanie Gr{\"a}fe, Jens Doose, and Joachim Burgd{\"o}rfer.
\newblock Quantum phase-space analysis of electronic rescattering dynamics in
  intense few-cycle laser fields.
\newblock \emph{Journal of Physics B: Atomic, Molecular and Optical Physics},
  45\penalty0 (5):\penalty0 055002, 2012.

\bibitem[Czirj{\'a}k et~al.(2000)Czirj{\'a}k, Kopold, Becker, Kleber, and
  Schleich]{czirjak2000ionizationwigner}
A~Czirj{\'a}k, R~Kopold, W~Becker, M~Kleber, and WP~Schleich.
\newblock The {Wigner} function for tunneling in a uniform static electric
  field.
\newblock \emph{Optics communications}, 179\penalty0 (1):\penalty0 29--38,
  2000.

\bibitem[Czirj{\'a}k et~al.(2013)Czirj{\'a}k, Majorosi, Kov{\'a}cs, and
  Benedict]{czirjak2013rescatterentanglement}
Attila Czirj{\'a}k, Szil{\'a}rd Majorosi, Judit Kov{\'a}cs, and Mih{\'a}ly~G
  Benedict.
\newblock Emergence of oscillations in quantum entanglement during
  rescattering.
\newblock \emph{Physica Scripta}, 2013\penalty0 (T153):\penalty0 014013, 2013.

\bibitem[Geltman(2011)]{geltman2011boundstatesdelta}
Sydney Geltman.
\newblock Bound states in delta function potentials.
\newblock \emph{Journal of Atomic, Molecular, and Optical Physics}, 2011, 2011.

\bibitem[Baumann et~al.(2015)Baumann, Kull, and
  Fraiman]{baumann2015wignerionization}
C~Baumann, H-J Kull, and GM~Fraiman.
\newblock Wigner representation of ionization and scattering in strong laser
  fields.
\newblock \emph{Physical Review A}, 92\penalty0 (6):\penalty0 063420, 2015.

\bibitem[Teeny et~al.(2016)Teeny, Yakaboylu, Bauke, and
  Keitel]{teeny2016ionizationtime}
Nicolas Teeny, Enderalp Yakaboylu, Heiko Bauke, and Christoph~H Keitel.
\newblock Ionization time and exit momentum in strong-field tunnel ionization.
\newblock \emph{Physical Review Letters}, 116\penalty0 (6):\penalty0 063003,
  2016.

\bibitem[Chomet et~al.(2019)Chomet, Sarkar, and
  de~Morisson~Faria]{chomet2019bridge}
H~Chomet, D~Sarkar, and C~Figueira de~Morisson~Faria.
\newblock Quantum bridges in phase space: interference and nonclassicality in
  strong-field enhanced ionisation.
\newblock \emph{New Journal of Physics}, 21\penalty0 (12):\penalty0 123004,
  2019.

\bibitem[Chomet and Figueira~de Morisson~Faria(2021)]{chomet2021phase}
H.~Chomet and C.~Figueira~de Morisson~Faria.
\newblock Attoscience in phase space.
\newblock \emph{Eur. Phys. J. D}, 75\penalty0 (201), 2021.

\bibitem[Chomet et~al.(2022)Chomet, Plesnik, Nicolae, Dunham, Gover, Weaving,
  and de~Morisson~Faria]{chomet2022machine}
H~Chomet, S~Plesnik, D~C Nicolae, J~Dunham, L~Gover, T~Weaving, and C~Figueira
  de~Morisson~Faria.
\newblock Controlling quantum effects in enhanced strong-field ionisation with
  machine-learning techniques.
\newblock \emph{Journal of Physics B: Atomic, Molecular and Optical Physics},
  55\penalty0 (24):\penalty0 245501, 2022.

\bibitem[Majorosi et~al.(2017)Majorosi, Benedict, and
  Czirj{\'a}k]{majorosi2017entanglement}
Szil{\'a}rd Majorosi, Mih{\'a}ly~G Benedict, and Attila Czirj{\'a}k.
\newblock Quantum entanglement in strong-field ionization.
\newblock \emph{Physical Review A}, 96\penalty0 (4):\penalty0 043412, 2017.

\bibitem[Hack et~al.(2021)Hack, Majorosi, Benedict, Varró, and
  Czirják]{hack2021interference}
Sz. Hack, Sz. Majorosi, M.~G. Benedict, S.~Varró, and A.~Czirják.
\newblock Quantum interference in strong-field ionization by a linearly
  polarized laser pulse and its relevance to tunnel exit time and momentum.
\newblock \emph{Physical Review A}, 104:\penalty0 L031102, 2021.

\bibitem[Sarkadi(2020)]{sarkadi2020nonseq}
L~Sarkadi.
\newblock Laser-induced nonsequential double ionization of helium: classical
  model calculations.
\newblock \emph{Journal of Physics B: Atomic, Molecular and Optical Physics},
  53\penalty0 (16):\penalty0 165401, 2020.

\bibitem[Truong et~al.(2022)Truong, Nguyen, Le, Dung, Tran, Vy, Anh-Tai, and
  Pham]{truong2022soft}
Thu~D.H. Truong, Hanh~H. Nguyen, Hieu~B. Le, Do~Hung Dung, H.-M. Tran,
  Nguyen~Duy Vy, Tran~Duong Anh-Tai, and Vinh~N.T. Pham.
\newblock Soft parameters in coulomb potential of noble atoms for nonsequential
  double ionization: Classical ensemble model and simulations.
\newblock \emph{Computer Physics Communications}, 276:\penalty0 108372, 2022.

\bibitem[Strelkov(2023)]{strelkov2023autoion}
V.~V. Strelkov.
\newblock Dark and bright autoionizing states in resonant high-order harmonic
  generation: Simulation via a one-dimensional helium model.
\newblock \emph{Phys. Rev. A}, 107:\penalty0 053506, 2023.

\bibitem[Harris(2023)]{harris2023ati}
A~L Harris.
\newblock Spectral phase effects in above threshold ionization.
\newblock \emph{Journal of Physics B: Atomic, Molecular and Optical Physics},
  56\penalty0 (9):\penalty0 095601, 2023.

\bibitem[Kocák and Schild(2020)]{kocak2020one}
Jakub Kocák and Axel Schild.
\newblock Many-electron effects of strong-field ionization described in an
  exact one-electron theory.
\newblock \emph{Phys. Rev. Res.}, 2:\penalty0 043365, 2020.

\bibitem[Ziems et~al.(2023)Ziems, Wollenhaupt, Gräfe, and
  Schubert]{ziems2023xuv}
Karl~Michael Ziems, Matthias Wollenhaupt, Stefanie Gräfe, and Alexander
  Schubert.
\newblock Attosecond ionization dynamics of modulated, few-cycle xuv pulses.
\newblock \emph{Journal of Physics B: Atomic, Molecular and Optical Physics},
  56\penalty0 (10):\penalty0 105602, 2023.

\bibitem[Lan et~al.(2023)Lan, Wang, Qiao, Zhou, Chen, Wang, Guo, and
  Yang]{lan2023twocolor}
Wendi Lan, Xinyu Wang, Yue Qiao, Shushan Zhou, Jigen Chen, Jun Wang, Fuming
  Guo, and Yujun Yang.
\newblock High-intensity harmonic generation with energy tunability produced by
  robust two-color linearly polarized laser fields.
\newblock \emph{Symmetry}, 15\penalty0 (3), 2023.
\newblock ISSN 2073-8994.

\bibitem[Wang et~al.(2021)Wang, Khan, Tian, Sun, and Jiang]{wang2021sch}
Shun Wang, Shahab~Ullah Khan, Xiao-Qing Tian, Hui-Bin Sun, and Wei-Chao Jiang.
\newblock Comparative study of photoionization of atomic hydrogen by solving
  the one- and three-dimensional time-dependent schrödinger equations*.
\newblock \emph{Chinese Physics B}, 30\penalty0 (8):\penalty0 083301, 2021.

\bibitem[Wang et~al.(2022)Wang, Li, Liu, Zhu, Jiao, and Liu]{wang2022mom}
Jun Wang, Gen-Liang Li, Xiaoyu Liu, Feng-Zheng Zhu, Li-Guang Jiao, and Aihua
  Liu.
\newblock Photoelectron momentum distribution of hydrogen atoms in a
  superintense ultrashort high-frequency pulse.
\newblock \emph{Frontiers in Physics}, 10, 2022.

\bibitem[Majorosi et~al.(2018)Majorosi, Benedict, and
  Czirják]{majorosi2018improved}
Szilárd Majorosi, Mihály~G. Benedict, and Attila Czirják.
\newblock Improved one-dimensional model potentials for strong-field
  simulations.
\newblock \emph{Phys. Rev. A}, 98:\penalty0 023401, 2018.

\bibitem[Majorosi and Czirj{\'a}k(2016)]{majorosi2016tdsesolve}
Szil{\'a}rd Majorosi and Attila Czirj{\'a}k.
\newblock Fourth order real space solver for the time-dependent
  {Schr{\"o}dinger} equation with singular {Coulomb} potential.
\newblock \emph{Computer Physics Communications}, 208:\penalty0 9--28, 2016.

\bibitem[Liu and Clark(1992)]{liu1992scgroundstate}
Wei-Chih Liu and C~W Clark.
\newblock Closed-form solutions of the schrodinger equation for a model
  one-dimensional hydrogen atom.
\newblock \emph{Journal of Physics B: Atomic, Molecular and Optical Physics},
  25\penalty0 (21):\penalty0 L517, nov 1992.

\bibitem[Bandrauk et~al.(2009)Bandrauk, Chelkowski, Diestler, Manz, and
  Yuan]{bandrauk2009tdsehydrogen}
AD~Bandrauk, S~Chelkowski, Dennis~J Diestler, J~Manz, and K-J Yuan.
\newblock Quantum simulation of high-order harmonic spectra of the hydrogen
  atom.
\newblock \emph{Physical Review A}, 79\penalty0 (2):\penalty0 023403, 2009.

\bibitem[Czirj{\'a}k et~al.(2012)Czirj{\'a}k, Majorosi, Kov{\'a}cs, and
  Benedict]{czirjak2012entanglementbuild}
Attila Czirj{\'a}k, Szil{\'a}rd Majorosi, Judit Kov{\'a}cs, and Mih{\'a}ly~G
  Benedict.
\newblock Build-up of quantum entanglement during rescattering.
\newblock In \emph{AIP Conference Proceedings}, volume 1462, pages 88--91. AIP,
  2012.

\bibitem[McPherson et~al.(1987)McPherson, Gibson, Jara, Johann, Luk, McIntyre,
  Boyer, and Rhodes]{McPherson_JOSAB_1987_HHG}
A~McPherson, G~Gibson, H~Jara, U~Johann, Ting~S Luk, IA~McIntyre, Keith Boyer,
  and Charles~K Rhodes.
\newblock Studies of multiphoton production of vacuum-ultraviolet radiation in
  the rare gases.
\newblock \emph{Journal of the Optical Society of America B}, 4\penalty0
  (4):\penalty0 595--601, 1987.

\bibitem[Harris et~al.(1993)Harris, Macklin, and
  H{\"a}nsch]{Harris_OptCom_1993_HHG_Atto}
SE~Harris, JJ~Macklin, and TW~H{\"a}nsch.
\newblock Atomic scale temporal structure inherent to high-order harmonic
  generation.
\newblock \emph{Optics communications}, 100\penalty0 (5-6):\penalty0 487--490,
  1993.

\bibitem[ELI()]{ELIlaser}
{ELI-ALPS} laser systems.
\newblock \url{https://www.eli-alps.hu/en/Users-2/SYLOS2-1;
  https://www.eli-alps.hu/en/Users-2/HR1-1}.
\newblock Accessed: 2024-01-15.

\bibitem[Ye et~al.(2020)Ye, Csizmadia, Oldal, Gopalakrishna, Füle, Filus,
  Nagyillés, Divéki, Grósz, Dumergue, Jójárt, Seres, Bengery, Zuba,
  Várallyay, Major, Frassetto, Devetta, Lucarelli, Lucchini, Moio, Stagira,
  Vozzi, Poletto, Nisoli, Charalambidis, Kahaly, Zaïr, and
  Varjú]{Ye2020beamline}
Peng Ye, Tamás Csizmadia, Lénárd~Gulyás Oldal, Harshitha~Nandiga
  Gopalakrishna, Miklós Füle, Zoltán Filus, Balázs Nagyillés, Zsolt
  Divéki, Tímea Grósz, Mathieu Dumergue, Péter Jójárt, Imre Seres, Zsolt
  Bengery, Viktor Zuba, Zoltán Várallyay, Balázs Major, Fabio Frassetto,
  Michele Devetta, Giacinto~Davide Lucarelli, Matteo Lucchini, Bruno Moio,
  Salvatore Stagira, Caterina Vozzi, Luca Poletto, Mauro Nisoli, Dimitris
  Charalambidis, Subhendu Kahaly, Amelle Zaïr, and Katalin Varjú.
\newblock Attosecond pulse generation at eli-alps 100 khz repetition rate
  beamline.
\newblock \emph{Journal of Physics B: Atomic, Molecular and Optical Physics},
  53\penalty0 (15):\penalty0 154004, jun 2020.

\bibitem[Ferray et~al.(1988)Ferray, L'Huillier, Li, Lompre, Mainfray, and
  Manus]{Ferray_JPhysB_1988_HHG}
M~Ferray, A~L'Huillier, XF~Li, LA~Lompre, G~Mainfray, and C~Manus.
\newblock Multiple-harmonic conversion of 1064 nm radiation in rare gases.
\newblock \emph{Journal of Physics B: Atomic, Molecular and Optical Physics},
  21\penalty0 (3):\penalty0 L31, 1988.

\bibitem[Gombk{\"o}t{\H{o}} et~al.(2016)Gombk{\"o}t{\H{o}}, Czirj{\'a}k,
  Varr{\'o}, and F{\"o}ldi]{gombkoto2016quantoptichhg}
{\'A}kos Gombk{\"o}t{\H{o}}, Attila Czirj{\'a}k, S{\'a}ndor Varr{\'o}, and
  P{\'e}ter F{\"o}ldi.
\newblock Quantum-optical model for the dynamics of high-order-harmonic
  generation.
\newblock \emph{Physical Review A}, 94\penalty0 (1):\penalty0 013853, 2016.

\bibitem[Farkas and T{\'o}th(1992)]{Farkas_PhysLettA_1992_AttoPulse}
Gy~Farkas and Cs~T{\'o}th.
\newblock Proposal for attosecond light pulse generation using laser induced
  multiple-harmonic conversion processes in rare gases.
\newblock \emph{Physics Letters A}, 168\penalty0 (5):\penalty0 447--450, 1992.

\bibitem[Paul et~al.(2001)Paul, Toma, Breger, Mullot, Aug{\'e}, Balcou, Muller,
  and Agostini]{Paul_Science_2001_Atto_pulsetrain}
PM~Paul, ES~Toma, P~Breger, Genevive Mullot, F~Aug{\'e}, Ph~Balcou, HG~Muller,
  and P~Agostini.
\newblock Observation of a train of attosecond pulses from high harmonic
  generation.
\newblock \emph{Science}, 292\penalty0 (5522):\penalty0 1689--1692, 2001.

\bibitem[Carrera et~al.(2006)Carrera, Tong, and Chu]{carrera2006attoxuv}
Juan~J Carrera, Xiao-Min Tong, and Shih-I Chu.
\newblock Creation and control of a single coherent attosecond {XUV} pulse by
  few-cycle intense laser pulses.
\newblock \emph{Physical review A}, 74\penalty0 (2):\penalty0 023404, 2006.

\bibitem[Sansone et~al.(2006)Sansone, Benedetti, Calegari, Vozzi, Avaldi,
  Flammini, Poletto, Villoresi, Altucci, Velotta,
  et~al.]{sansone2006attosecondisolated}
Giuseppe Sansone, E~Benedetti, Francesca Calegari, Caterina Vozzi, Lorenzo
  Avaldi, Roberto Flammini, Luca Poletto, P~Villoresi, C~Altucci, R~Velotta,
  et~al.
\newblock Isolated single-cycle attosecond pulses.
\newblock \emph{Science}, 314\penalty0 (5798):\penalty0 443--446, 2006.

\end{thebibliography}

\end{document}